\documentclass[iop,apj]{emulateapj}

\newcommand{\ax}{SN~2008ax}
\newcommand{\synow}{\textsc{SYNOW}}
\newcommand{\hal}{H$\alpha$}
\newcommand{\hbeta}{H$\beta$}
\newcommand{\host}{NGC~4490}
\newcommand{\kms}{km~s$^{-1}$}
\newcommand{\pap}{Li et al., in prep.}
\newcommand{\citepeg}[1]{\citep[{e.g.,}][]{#1}}
\newcommand{\qrsp}{$q_{\rm{RSP}}$}
\newcommand{\ursp}{$u_{\rm{RSP}}$}
\newcommand{\ebv}{$E(B-V)$}

\begin{document}

\shorttitle{The Transitional Supernova 2008ax}
\shortauthors{Chornock et al.}

\title{The Transitional Stripped-Envelope SN 2008ax: Spectral
  Evolution and Evidence for Large Asphericity}

\author{R. Chornock\altaffilmark{1,2,3},
A. V. Filippenko\altaffilmark{1},
W. Li\altaffilmark{1},
G. H. Marion\altaffilmark{4,5},
R. J. Foley\altaffilmark{2,6},
M. Modjaz\altaffilmark{1,7},
M. Rafelski\altaffilmark{8},
G. D. Becker\altaffilmark{9},
W. H. de Vries\altaffilmark{10,11},
P. Garnavich\altaffilmark{12},
R. A. Jorgenson\altaffilmark{13},
D. K. Lynch\altaffilmark{14},
A. L. Malec\altaffilmark{15},
E. C. Moran\altaffilmark{16},
M. T. Murphy\altaffilmark{15},
R. J. Rudy\altaffilmark{14},
R. W. Russell\altaffilmark{14},
J. M. Silverman\altaffilmark{1},
T. N. Steele\altaffilmark{1},
A. Stockton\altaffilmark{17},
A. M. Wolfe\altaffilmark{8},
and C. E. Woodward\altaffilmark{18}
}

\altaffiltext{1}{Department of Astronomy, University of California,
                 Berkeley, CA 94720-3411, USA.}
\altaffiltext{2}{Harvard-Smithsonian Center for Astrophysics, 60
  Garden Street, Cambridge, MA 02138, USA.}
\altaffiltext{3}{\texttt{rchornock@cfa.harvard.edu}.}
\altaffiltext{4}{The George P. \& Cynthia W. Mitchell Institute for
  Fundamental Physics and Astronomy, Texas A\&M University, Department
  of Physics and 
  Astronomy, 4242 TAMU, College Station, TX 77843-4242, USA.}
\altaffiltext{5}{University of Texas, Department of Astronomy, Austin,
  TX 78712, USA.} 
\altaffiltext{6}{Clay Fellow.}
\altaffiltext{7}{Miller Fellow.}
\altaffiltext{8}{Department of Physics, and Center for Astrophysics
  and Space Sciences, University of California, San Diego, 9500 Gilman
  Drive, La Jolla, CA 92093-0424, USA.}
\altaffiltext{9}{Fellow, Kavli Institute for Cosmology and Institute
  of Astronomy, Madingley Road, Cambridge CB3 0HA, UK.}
\altaffiltext{10}{University of California, Department of Physics, 1
  Shields Ave, Davis, CA 95616, USA.}
\altaffiltext{11}{Institute for Geophysics and Planetary Physics,
  LLNL, L-413, 7000 East Avenue, Livermore, CA 94550, USA.} 
\altaffiltext{12}{Physics Department, University of Notre Dame, Notre
  Dame, IN 46556, USA.} 
\altaffiltext{13}{Institute of Astronomy, University of Cambridge,
  Madingley Road, Cambridge, CB3 0HA, UK.} 
\altaffiltext{14}{The Aerospace Corporation, Mail Stop M2-266, PO Box
  92957, Los Angeles, CA 90009-2957, USA.} 
\altaffiltext{15}{Centre for Astrophysics and Supercomputing,
  Swinburne University of Technology, Mail H39, PO Box 218, Victoria
  3122, Australia.}
\altaffiltext{16}{Astronomy Department, Wesleyan University,
  Middletown, CT 06459, USA.} 
\altaffiltext{17}{Institute for Astronomy, University of Hawaii, 2680
  Woodlawn Drive, Honolulu, HI 96822, USA.} 
\altaffiltext{18}{Department of Astronomy, University of Minnesota,
  116 Church Street, S. E., Minneapolis, MN 55455, USA.}

\begin{abstract}
Supernova (SN) 2008ax in NGC 4490 was discovered within hours after shock
breakout, presenting the rare opportunity to study a core-collapse SN
beginning with the initial envelope-cooling phase immediately following
shock breakout.  We present an extensive sequence of optical and
near-infrared spectra, as well as three epochs of optical
spectropolarimetry.  Our initial spectra, taken two days after shock
breakout, are dominated by hydrogen Balmer lines at high velocity.
However, by maximum light, \ion{He}{1} lines dominated the optical and
near-infrared spectra, which closely resembled those of normal Type~Ib
supernovae (SNe Ib) such as SN 1999ex.  This spectroscopic transition
defines Type~IIb supernovae, but the strong similarity of SN 2008ax
to normal SNe Ib beginning near maximum light, including an absorption
feature near 6270~\AA\ due to \hal\ at high velocities, suggests that
many objects classified as SNe Ib in the literature may have ejected
similar amounts of hydrogen as SN 2008ax, roughly a few $\times$~0.01
M$_{\odot}$.  Only the unusually early discovery of \ax\ allowed us to
observe the spectroscopic signatures of the hydrogen-rich outer
ejecta.  Early-time spectropolarimetry (6 and 9 days after shock
breakout) revealed strong line
polarization modulations of 3.4\% across \hal, indicating the presence
of large asphericities in the outer ejecta and possibly that the
spectrum of \ax\ could be dependent on the viewing angle. 
After removal of interstellar polarization, the continuum shares a
common polarization angle with the hydrogen, helium, and oxygen lines,
while the calcium and iron absorptions are oriented at different
angles.  This is clear evidence of deviations from axisymmetry even in
the outer ejecta.  Intrinsic continuum polarization of 0.64\% only
nine days after shock breakout shows that the outer layers of the
ejecta were quite aspherical.  A single epoch of late-time
spectropolarimetry, as well as the shapes of the nebular line
profiles, demonstrate that asphericities extended from the outermost
layers all the way down to the center of this core-collapse SN.  SN
2008ax may in some ways be an extragalactic analog of the explosion
giving rise to Cassiopeia~A, which has recently been determined to be
a remnant of a SN IIb.
\end{abstract}

\section{Introduction}

Massive stars explode at the end of their lives as core-collapse
supernovae (SNe), which we observe and classify spectroscopically on
the basis of their optical spectra near maximum light (see Filippenko
1997 for a review).  These phenomenological spectroscopic classes are
related to the state of the progenitor star at the time of explosion
\citepeg{heger03}.  Supernovae coming from stars that retain most of
their massive hydrogen envelopes  
until the time of explosion are classified as Type II, while those
lacking hydrogen due to either stellar winds or interaction with a
companion star are classified as Types Ib and Ic.  Progenitors with
only small amounts of hydrogen remaining in their outer envelopes
explode as SNe IIb, a classification named because
their spectra transition from Type II at early times to Type Ib at
late times. 

SNe~IIb are up to $\sim$11\% of the core-collapse SN population in a
distance-limited sample \citep{li11,smith11}.  Classical
SNe~Ib are a further 7\% of the core-collapse supernova sample,
although some of those objects (such as SN 1999dn; Branch et al. 2002)
likely had weak \hal\ absorption in their early-time spectra
from a small residual hydrogen envelope.  We will use the catch-all
term ``stripped-envelope supernovae'' \citep{cloch96} to refer to SNe
from massive stars lacking the thick hydrogen envelopes of typical SNe
II.

The prototypical SN~IIb, SN 1993J,
occurred in M81 and was one of the brightest SNe of the 20th century.
Its proximity and late-time brightness at all wavelengths due to
strong interaction with circumstellar material (CSM) enabled numerous
detailed studies that make it 
the second-best studied core-collapse SN after SN 1987A in the Large
Magellanic Cloud \citep{wf96}.  Furthermore, one of the youngest and
best-studied supernova remnants in the Galaxy, Cassiopeia A (Cas A),
was recently revealed to have been a SN IIb on the basis of
spectroscopy of scattered-light echoes from the supernova
\citep{krause08}.

We gained the opportunity to study another nearby stripped-envelope SN
in great detail when \citet{mostardi08} announced the discovery of a
possible supernova at unfiltered magnitude 16.1 in the nearby ($d
\approx 9.6$~Mpc; Pastorello et al. 2008 [P08 hereafter]) galaxy NGC
4490 on 2008 March 3.45 (UT dates are used throughout this paper).
The following night, the new object was spectroscopically confirmed as
a peculiar SN II \citep{blondin_cbet}.  The spectroscopic
classification continued to evolve as the supernova aged, with
\citet{me08axcbet} identifying the incipient \ion{He}{1} lines
diagnostic of SNe IIb by March 9.6 and \citet{marion08} using
near-infrared (NIR) spectra of \ax\ on March 14.5 to classify it as
typical of SNe Ib.

Fortuitously, \citet{arbour_cbet} had imaged NGC 4490 on March 3.19
and did not detect \ax\ to a limiting unfiltered magnitude of 19.5
(P08). The supernova must have risen by at least 3.4 mag in the
six hours before the first detection by \citet{mostardi08}, setting a
strict limit on the time of shock breakout.  Throughout this paper
we will assume a time for shock breakout of \ax\ of 2008 March 3.32,
halfway between the non-detection by \citet{arbour_cbet} and the
discovery by \citet{mostardi08}.  This unusually early discovery
enabled rapid follow-up observations at all wavelengths.

Observations of SNe at very early times are obtained only rarely, but
they are important because they reflect the state of the outermost
layers of the progenitor star, which are otherwise difficult to study,
and provide a crucial link between studies of core-collapse SNe and
their progenitors.  The ultraviolet (UV) and optical light 
curves presented by \citet{roming09} show the initial phase of rapid
cooling of the stellar envelope after shock breakout, a short-lived
phenomenon occurring immediately after explosion that has only been
observed in a handful of core-collapse SNe.  The proximity of the host
galaxy enabled \citet{li08} and \citet{crockett08} to identify a
source at the position of \ax\ in pre-explosion \emph{Hubble Space
  Telescope} images.  Unfortunately, the apparent progenitor appears
to be blended with other nearby sources, making the interpretation
ambiguous. \citet{crockett08} found that the progenitor could be
modeled as either a massive star that had lost most of its hydrogen
envelope in a Wolf-Rayet (WR) phase or as an interacting binary star.

We conducted an extensive spectroscopic campaign to observe \ax,
starting two days after shock breakout and extending into the nebular
phase.  Our earliest spectra are sensitive to the
composition of the outermost layers of the progenitor, while the
latest ones probe the inner core.  In addition, five epochs of NIR
spectra were taken over the first two months after explosion, starting
10 days after shock breakout.  The NIR spectra of stripped-envelope
SNe are less well-studied than those at optical wavelengths, but they
provide a different window into the explosion with strong lines from
light and intermediate-mass elements that are blended or obscured at
optical wavelengths.  The importance of
\ax\ also led us to obtain three epochs of optical spectropolarimetry.
Our first two observations were obtained within 9 days of shock
breakout and allow us to study the geometry of the outer ejecta, while
our final epoch extends our study of the geometry to the early nebular
phase.

Optical photometry of \ax\ has been presented by P08, as well as a
series of seven optical spectra starting 10 days after shock
breakout and a single NIR spectrum obtained after maximum light.
Additional photometry and preliminary modeling of the light curve were
presented by \citet{tsv09}.  Late-time optical spectra of \ax\ were
studied in detail by \citet{mili09}.  \citet{taub11} have also
presented a large optical and NIR dataset.  \citet{roming09} presented
UV and optical light curves, as well as X-ray and radio detections,
enabling study of pre-SN mass loss of the progenitor.
\citet{roming09} also showed two optical spectra taken near maximum
light.

In this paper, we add to the previous studies of \ax\ by presenting
our large and well-sampled spectroscopic dataset.  The observations
are described in \S2 and the detailed optical through NIR spectral
evolution is discussed in \S3.  A highlight of our study are the three
epochs of optical spectropolarimetry showing large line polarizations
beginning shortly after explosion, which are analyzed in \S4.
Implications of these observations are discussed in \S5.

\section{Observations\label{ax_obs}}

\subsection{Optical Spectroscopy}

\begin{figure*}
\plotone{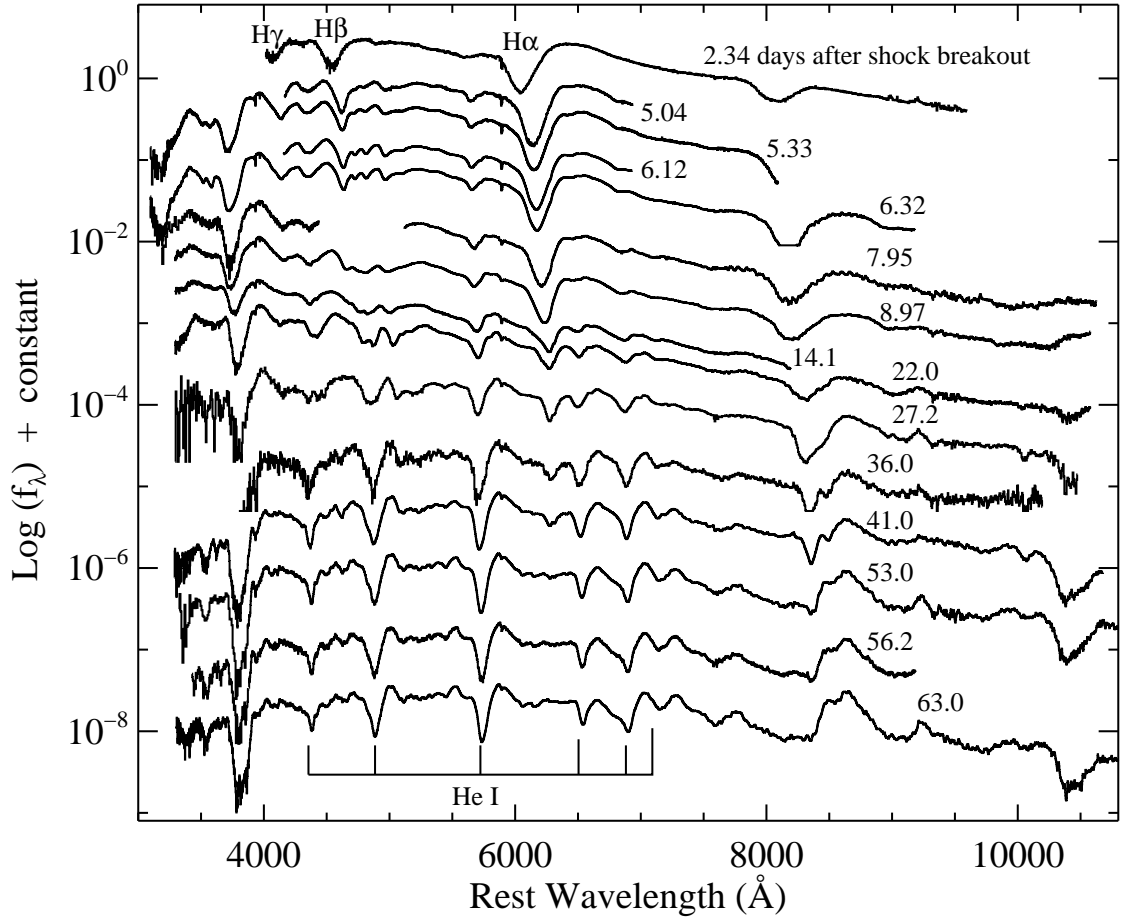}
\caption[Optical spectral sequence of \ax\ during the photospheric
  phase.]{Optical spectral sequence of \ax\ during the photospheric
  phase.  Spectra are labeled with their ages in days relative to the
  assumed
  shock-breakout date of 2008 March 3.32.  Maximum light in the
  $B$ band was 19 days later (March 22.2; P08).  The earliest spectrum
  shows 
  prominent absorption lines from the hydrogen Balmer series.  The He
  I lines are visible shortly after explosion and gradually come to
  dominate the spectrum.  By two months after explosion, the H$\alpha$
  absorption line disappeared. 
}
\label{ax_allplot}
\end{figure*}

We obtained a well-sampled sequence of low-resolution optical spectra
of SN 2008ax starting two days after discovery and extending into the early
nebular phase.  Spectra were obtained using the Kast spectrograph on
the Lick 3-m telescope \citep{ms93}, the Low Resolution Imaging
Spectrometer \citep{oke95} on the Keck I 10-m telescope, and the
Echellette Spectrograph and Imager (ESI; Sheinis et al. 2002) on the
Keck II 10-m telescope.
Two-dimensional image reduction and spectral extraction were performed
using standard tasks in IRAF\footnote{IRAF is distributed by the
  National Optical Astronomy Observatories, 
    which are operated by the Association of Universities for Research
    in Astronomy, Inc., under cooperative agreement with the National
    Science Foundation.}.  Flux calibration and removal of telluric
absorption features were performed using our own IDL routines
\citep{ma00b}.  The early-time spectra are shown in 
Figure~\ref{ax_allplot}.  A log with the details of all the observations
is given in Table~\ref{ax_specobstab}.  Hereafter, we shall refer to
the dates of observation as ``day X,'' where X represents the number
of days after shock breakout, assumed to be 2008 March 3.32.  P08
determined the time of maximum light in the $B$ band to be 19 days
later (2008 March $22.2 \pm 0.5$).

\begin{deluxetable*}{cccccccc}
\tabletypesize{\scriptsize}
\tablecaption{Log of Low-Resolution Optical Spectroscopy}
\tablehead{\colhead{Age\tablenotemark{a}} & \colhead{UT
    Date\tablenotemark{b}} & 
\colhead{Instrument\tablenotemark{c}} &
\colhead{Range} & \colhead{Exposure Time} & \colhead{Seeing} &
\colhead{Airmass} & \colhead{Clouds?} \\
\colhead{(days)} & (YYYY-MM-DD.DDD) &  & \colhead{(\AA)}
  & \colhead{(s)} & \colhead{($\arcsec$)} &  & \colhead{(y/n)} }
\startdata
2.34 & 2008-03-05.658 & ESI & 4000$-$9600 & 300 & 0.9 & 1.6 & n \\
5.04 & 2008-03-08.360 & Kast & 4170$-$6950 & 2700 & 2.3 & 1.0 & n \\
5.33 & 2008-03-08.650 & LRIS & 3100$-$8110 & 200 & 0.8 & 1.6 & n \\
6.12 & 2008-03-09.437 & Kast(p) & 4170$-$6940 & 4$\times$3000\tablenotemark{d} & 2.1 & 1.1 & n \\
6.32 & 2008-03-09.638 & LRIS & 3100$-$9200 & 560/480\tablenotemark{e} & 0.7 & 1.6 & n \\
7.96 & 2008-03-11.283 & Kast & 3110$-$4450, & 1200 & 2.2 & 1.2 & y \\
 & & & 5130$-$10650 & & & & \\
8.97 & 2008-03-12.292 & Kast(p) & 4390$-$9880 & 4$\times$3000\tablenotemark{d} & 2.0 & 1.2 & y \\
9.07 & 2008-03-12.393 & Kast & 3300$-$10600 & 1200 & 3.0 & 1.0 & y \\
14.1 & 2008-03-17.423 & Kast & 3300$-$8210 & 600 & 2.3 & 1.0 & n \\
22.0 & 2008-03-25.348 & Kast & 3300$-$10600 & 400 & 2.7 & 1.0 & y \\
27.2 & 2008-03-30.548 & Kast & 3300$-$10500 & 400 & 3.4 & 1.8 & y \\
36.0 & 2008-04-08.312 & Kast & 3300$-$10800 & 600/700\tablenotemark{e} & 2.4 & 1.0 & y \\
41.0 & 2008-04-13.314 & Kast & 3300$-$10700 & 1600 & 2.0 & 1.0 & y \\
53.0 & 2008-04-25.351 & Kast & 3300$-$10800 & 1800 & 2.3 & 1.1 & y \\
56.2 & 2008-04-28.558 & LRIS & 3000$-$9200 & 500 & 1.9 & 2.3 & y \\
63.0 & 2008-05-05.331 & Kast & 3300$-$10800 & 1800 & 2.0 & 1.1 & y \\
85.1 & 2008-05-27.382 & Kast & 3300$-$10800 & 1500/1562\tablenotemark{e} & 1.9 & 1.7 & y \\
99.0 & 2008-06-10.310 & Kast(p) & 4580$-$10060 & 4$\times$2400\tablenotemark{d} & 2.5 & 1.6 & n \\
118 & 2008-06-29.249 & Kast & 3300$-$10750 & 1800 & 2.1 & 1.3 & y \\
158 & 2008-08-08.914 & Kast & 3540$-$10750 & 2100 & 2.5 & 1.9 & n \\
\enddata
\tablenotetext{a}{Age in days relative to the assumed shock-breakout
  date of 2008 March 3.32.}
\tablenotetext{b}{Midpoint of observation.}
\tablenotetext{c}{ESI = Echellette Spectrograph and Imager on Keck-II
  10-m telescope. Kast = Kast spectrograph on Lick 3-m telescope; (p)
  if used in polarimetry mode. LRIS
  = Low Resolution Imaging Spectrometer on Keck-I 10-m telescope.} 
\tablenotetext{d}{Exposure time was (for example) 3000~s at each of the
  four waveplate rotation angles, for a total of 12,000~s.}
\tablenotetext{e}{Total exposure times on blue/red sides of spectrograph.}
\label{ax_specobstab}
\end{deluxetable*}

\begin{figure}
\plotone{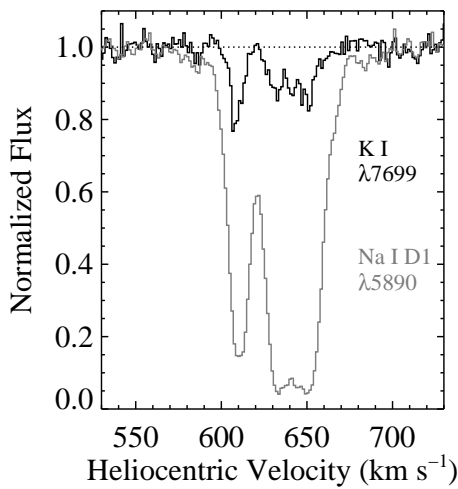}
\caption[Na~I and K~I lines in NGC 4490]{\ion{Na}{1} and
  \ion{K}{1} lines from the interstellar medium of NGC 4490 in the
  HIRES spectrum of \ax.  The \ion{Na}{1} absorption is clearly
  saturated and is not a reliable tracer of the total column along the
  line of sight to \ax.  We adopt 630 \kms\ as the radial velocity of
  \ax.
}
\label{ax_naiplot}
\end{figure}

In addition, we obtained high-resolution spectra of \ax\ using
the High Resolution Echelle Spectrometer (HIRES; Vogt et al. 1994) on
Keck I on 2008 March 24 and 25, with exposure times of 1200~s on each
night.  Numerous absorption features from the interstellar medium
(ISM) of \host\ are present, including diffuse interstellar bands
(DIBs; Herbig 1975), which we discuss further in the Appendix.  
The \ion{Na}{1} D1 $\lambda$5890 and
\ion{K}{1} $\lambda$7699 absorption-line profiles are shown in
Figure~\ref{ax_naiplot}.  Both lines show absorption from multiple
components over the velocity range of 590$-$670 \kms, so we adopt 630
\kms\ as the radial velocity of \ax.  This redshift has been removed
from all other spectra in the plots in this paper.

These lines can also be used to estimate the host-galaxy contribution 
to the reddening to \ax, which dominates over the Galactic
component ($E(B-V) = 0.02$~mag; Schlegel et al. 1998).  However, the
\ion{Na}{1} lines which are usually used to estimate the reddening in
SN studies have multiple saturated components along this line of sight
and therefore do not provide a reliable reddening estimate, so instead
we use the \ion{K}{1} $\lambda$7699 line.  The total equivalent 
width (EW) of the \ion{K}{1} $\lambda$7699 absorption over the
velocity range of 590$-$670 \kms\ is $0.142 \pm 0.006$~\AA, which
corresponds to a predicted reddening of $E(B-V) = 0.54$~mag using the 
relationship of \citet{munz}.  This value, combined with a detailed
comparison of the colors of \ax\ to those of other core-collapse SNe
(\pap), leads us to adopt a value for the reddening of the SN of
$E(B-V) = 0.5$~mag.  All spectra in this paper have been corrected for
this 
value of the reddening (assuming that $R_V = A_V / E(B-V) = 3.1$).
This value is higher than that used by P08 ($E(B-V) = 0.3$~mag) and
\citet{crockett08}, which was estimated using low-resolution
spectra of the saturated \ion{Na}{1} line.  Our spectroscopic
analysis is insensitive to the exact value of the adopted reddening.

\subsection{Near-Infrared Spectroscopy}

\begin{figure}
\plotone{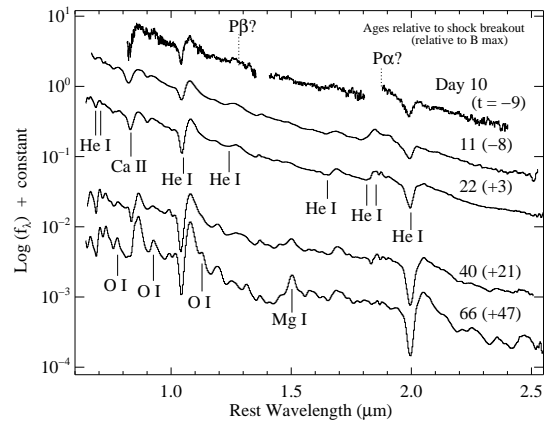}
\caption[NIR spectra of \ax.]{NIR spectra of \ax.  The spectra are
  labeled by dates relative to shock breakout and relative to maximum
  light in $B$ (in parentheses).  The spectra obtained prior to
  maximum light are
  smooth, with few strong lines.  Dotted lines mark the rest
  wavelengths of P$\alpha$ and P$\beta$, the strongest
  expected lines of hydrogen, but there are no clear associated
  emission peaks.  The expected 
  locations of absorption minima from \ion{He}{1} lines at 0.7065,
  0.7281, 1.0830, 1.2787, 1.7002, 1.8689, 1.9089, and 2.0581 $\mu$m,
  blueshifted by 9000 \kms, are labeled in the near-maximum-light
  spectrum (day 22).  After maximum light, blends of features from
  neutral and singly ionized intermediate-mass elements become more
  prominent.  
}
\label{ax_irplot}
\end{figure}

We observed \ax\ on five epochs (2008 Mar. 13.51, 14.52, 25.44,
Apr. 12.48, and May 8.30; days 10, 11, 22, 40, and 66) using the SpeX
instrument on the 3-m NASA
Infrared Telescope Facility \citep{specxref}.  Our first observation
was taken in cross-dispersed mode (SXD; $R = \lambda/\Delta\lambda
\approx 1200$), but the remaining observations used the 0$\farcs$5
slit in low-resolution prism mode (LRS; $R \approx 200$) to cover the
wavelength range 0.7$-$2.4~$\mu$m.  The reductions were performed
using the Spextool package \citep{cushing04} and corrections for
telluric absorption were made following the method of \citet{vacca03}.
A thorough discussion of the data acquisition and reduction process in
the context of supernova observations was presented by \citet{mar09}.
The NIR spectra of \ax\ are shown in Figure~\ref{ax_irplot}. 

\subsection{Spectropolarimetry}

We obtained three epochs of optical spectropolarimetry using Kast at
the Lick 3-m telescope, twice at early phases (days 6 and 9) and once
in the early nebular phase (day 99).  The data were reduced in a
standard manner, following the method described by \citet{mrg88} and
implemented by \citet{le01}.  All of the spectropolarimetry was taken
using only the red side of Kast.  On days 9 and 99 we used a GG455
order-blocking filter, but our spectral range extended to wavelengths
greater than 9000~\AA, where some second-order light is likely
present.  Our analysis does not rely on data at these wavelengths,
although the red continuum of \ax\ due to its high extinction implies
that contamination should be small.

We observed polarimetric standards from \citet{sch92} to set the zero
point of the half-wave plate retarder and
low-polarization standards from the literature \citep{sch92} to verify
that instrumental polarization was negligible ($P < 0.1$\%).  On day
99, an instrumental failure unrelated to the polarimeter unit prevented
us from obtaining polarization standards, so we used the
angle-correction curve from day 9.  Past experience has shown that the
zero point of the Kast waveplate is usually stable, with only small
drifts over time.  The day 6 and 9 zero points agree to within
$1\degr$ of data taken as far in the past as at least 2007 May 10
\citep{meIIP}. Our next available spectropolarimetry
data after the day 99 dataset were taken on 2008 Dec. 31 and the zero
point had shifted by 2$\fdg$7 relative to day 9.  Therefore, we
believe our lack of observations of standards on day 99 results in
little uncertainty in our calibrations.

\section{Spectral Evolution\label{ax_spectra}}

The optical spectral evolution of \ax\ in the photospheric phase can
be seen in Figure~\ref{ax_allplot}.  The sequence shows an obvious trend
as the hydrogen Balmer lines decrease in strength and shift to lower
velocities over time while the \ion{He}{1} lines strengthen and
dominate the spectra after maximum light (see also P08).

\subsection{Day 2 Spectrum}

\begin{figure}
\plotone{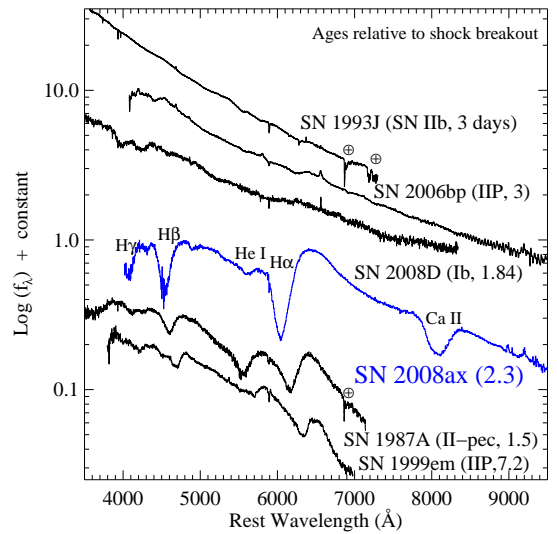}
\caption[Very early spectra of several core-collapse SNe compared to
 that of \ax.]{Very early spectra of several 
  core-collapse SNe of a variety of spectral types compared to \ax. 
 The spectra are labeled with their ages relative to assumed times of
 shock breakout.  The moderate-resolution ESI spectrum of \ax\ has
 been rebinned for display purposes.  Spectra of other objects were
 collected from the literature (SN 1987A: Menzies et al. 1987; SN
 1993J: Filippenko  et al. 1993; SN
 1999em: Leonard et al. 2002; SN 2006bp: Quimby et al. 2007; SN 2008D:
 Modjaz et al. 2009).  Uncorrected telluric absorptions in the
 literature spectra are marked with a $\earth$ symbol.  All spectra in
 this and subsequent comparison plots were corrected for the reddening
 values listed in Table~\ref{ax_redtab}, assuming that $R_V = 3.1$.
}
\label{ax_earlyplot}
\end{figure}

Our first spectrum, taken just 2.34 days after shock breakout, is one
of the earliest spectra of a SN ever taken.  We compare it in
Figure~\ref{ax_earlyplot} to the earliest spectra of other core-collapse
SNe of various spectral types.  Throughout this paper, we correct
the reference SN spectra for reddening to facilitate comparison, using
the adopted values listed in Table~\ref{ax_redtab}.  In a few cases,
good reddening estimates do not exist, so we do not list those objects
in Table~\ref{ax_redtab} or correct their spectra for reddening.

SN spectral types are defined using optical spectra near maximum light
\citepeg{fil97}, but at early times the spectra do not correspond in a
simple manner to classes based on their appearances at later times.
The diversity of these spectra is 
related to the state of the progenitor star at the time of explosion
and to the presence or absence of CSM.  The SN IIb 1993J and the SN IIP
2006bp both have blue spectra with relatively weak spectral features.
Both also showed transient narrow emission features from the CSM that
was overrun by the SN ejecta within a few days
\citep{ben94,ga94,quimby06bp}.
The SNe IIP 1999em and 2006bp resembled each other at 
later epochs, but in these early-time spectra the Balmer lines clearly 
have different strengths.  Similarly, the early spectra of SNe 2008D
\citep{soderberg08,modjaz09,mazzali08,malesani09} and
2008ax are very different, with SN 2008ax showing strong Balmer lines
from its very thin outer H envelope, but by maximum light the two
SNe showed a close resemblance (see below). 

\begin{deluxetable}{lcl}
\tablecaption{Adopted Reddenings for Comparison SN Spectra}
\tablehead{\colhead{Object} & \colhead{\ebv} &
 \colhead{Note about source}\\
 & (mag) & }
\startdata
SN~1987A & 0.17 & \citet{ws90} \\
SN~1987K & 0.5 & To match continuum shape with \ax\ \\
SN~1990B & 0.85 & \citet{cloch90b} \\
SN~1993J & 0.3 & ``Moderate'' value from \citet{rich94} \\
SN~1999em & 0.1 & \citet{baron00} \\
SN~1999ex & 0.5 & To match observed colors with \ax\ \\
SN~2000H & 0.16 & \citet{ben00} \\
SN~2005bf & 0.045 & \citet{folatelli06} \\
SN~2006bp & 0.4 & \citet{dessart06bp} \\
SN~2008D & 0.6 & \citet{modjaz09} \\
\enddata
\label{ax_redtab}
\end{deluxetable}

The most distinctive aspect of \ax\ in Figure~\ref{ax_earlyplot} is the
series of extremely strong absorptions from Balmer transitions and
the \ion{Ca}{2} NIR triplet with very high expansion velocities.  The
absorptions are clearly deeper than in any of the other objects.
By comparison, SNe~1993J and 2008D show only weak features.
\citet{modjaz09} attributed the strongest features in this SN 2008D
spectrum to \ion{C}{3} and \ion{N}{3}, two species indicative of
higher temperatures than the low-ionization H and \ion{Ca}{2} lines
present in the spectrum of \ax. The weak features in the earliest
spectra of SN~1993J are not well understood \citep{baron95}, but
\hal\ is likely present.

The large velocities of the Balmer lines in our day 2.34 \ax\ spectrum,
even in comparison to 
SN~1987A, are also evident in Figure~\ref{ax_earlyplot}.  The flux minimum
of \hal\ is blueshifted by 24,600 \kms\ and the blue wing of absorption
extends outward to greater than 33,000 \kms, with the precise value
slightly uncertain due to the difficulty in defining a continuum with
such broad lines.  \citet{blondin_cbet}
reported that on the previous night (day 1.05 after shock breakout)
the minimum of \hal\ was blueshifted by 26,600 \kms.  Similarly, the
\ion{Ca}{2} NIR triplet absorption minimum was near 8075~\AA, with the
blue wing extending outward to 7800~\AA\ (blueshifts of 18,100 and
28,500 \kms\ relative to the $gf$-weighted line centroid of 8579~\AA,
respectively).  The 
\hal\ profile in the earliest SN 1987A spectrum from
\citet{menzies87}, taken 1.5 days after explosion (and shown in
Figure~\ref{ax_earlyplot}), has a flux minimum 
blueshifted by 18,300 \kms.  By comparison, the \hal\ minimum of the
normal SN IIP 2006bp was blueshifted by only 15,400 \kms\ three days
after shock breakout \citep{quimby06bp}\footnote{We have adopted a
  date for the shock breakout of SN~2006bp of 2006 April 7.9,
  following the model fits of \citet{dessart06bp}, which is a day
  earlier than assumed by \citet{quimby06bp}.}.  The broad-lined SN
IIb 2003bg did have \hal\ absorption velocities near 20,000
\kms\ in the earliest spectra \citep{hamuy09,mazzali09}.

\citet{matznermckee} investigated the
effects of supernova shock waves on the outer envelopes of their
progenitors and produced expressions for the maximum ejecta
velocities.  For typical core-collapse SN parameters (e.g.,
$M_{\rm{ejecta}} = 10$~M$_{\odot}$, $E = 10^{51}$ ergs), they find
that the outer layers of SN progenitors
with extended convective envelopes, such as the red supergiant
progenitors of SNe~IIP \citepeg{smartt09}, are only
accelerated up to velocities of $\sim$13,000 \kms.  The outer layers
of compact progenitors with radiative envelopes, such as blue
supergiants or helium stars, can be accelerated up to $\sim$33,000
\kms\ with the same SN parameters.  The very high ejecta velocities of
\ax\ are evidence of a compact progenitor star that is independent of
the direct analysis of pre-explosion photometry of the progenitor
\citep{crockett08} or the early-time light curve \citep{roming09}.

If we adopt the derived values for the explosion energy and mass of
the ejecta ($M_{\rm{ejecta}} = 2.9$~M$_{\odot}$, $E = 0.5 \times
10^{51}$ ergs) from the analysis of \citet{roming09} and keep the
fiducial values for the other parameters, the relationships of
\citet{matznermckee} predict that the highest expansion velocities for
\ax\ should be $\sim$37,000 \kms, consistent with the optical
observations.  However, \citet{mv09} used Very Long Baseline
Interferometry at 23 GHz to measure an expansion velocity for the SN
shock of ($5.2 \pm 1.3$) $\times$ $10^4$~\kms, indicating that our
optical observations are (unsurprisingly) not sensitive to the very
highest velocity material.  If instead we had adopted the explosion
parameters from the light-curve modeling of \citet{tsv09}
($M_{\rm{ejecta}} = 3.8$~M$_{\odot}$, $E = 1.5 \times 10^{51}$ ergs, $R_*$ =
600~R$_{\odot}$), we would get a maximum velocity of 28,300 \kms.
This value is clearly an underestimate due to the overly large
progenitor radius, which also results in their model overpredicting
the observed luminosity of the envelope cooling emission in the early
light curve.

As mentioned above, both the SN IIb 1993J and the SN IIP 2006bp
exhibited narrow emission lines in their earliest spectra indicative
of CSM ionized by both the supernova and the CSM interaction shock.
SN~2006bp had both the expected narrow Balmer lines as well as
emission from higher-ionization material represented by \ion{He}{2}
$\lambda\lambda$4200, 4686 and \ion{C}{4} $\lambda$5805
\citep{quimby06bp}.  The stronger CSM interaction component of
SN~1993J created high-temperature gas in the shock, which in turn
resulted in sufficient ionization in the unshocked CSM to produce
emission from [\ion{Fe}{10}] $\lambda$6375 and [\ion{Fe}{14}]
$\lambda$5303 \citep{ben94,ga94}.  

Our moderate-resolution ESI spectrum from day 2.34
afforded an excellent opportunity to search for such features in \ax.
However, none are detected. 
We used the noise level in the spectrum to set 3$\sigma$ limits on
the EWs of any unresolved narrow emission lines of
0.05--0.15~\AA, dependent on wavelength.  These limits are
significantly lower than the observed EWs of the features in SN 2006bp
(0.7--3.1 \AA; Quimby et al. 2007) or SN 1993J (0.11--0.3 \AA;
Benetti et al. 1994) at similar epochs.  When scaled to the
photometry provided by P08, the upper limits in flux at
($\lambda$4686, $\lambda$6375) are (1.5, 0.8) $\times$ 10$^{-16}$
ergs~cm$^{-2}$~s$^{-1}$, which correspond to dereddened luminosities
of $\lesssim$(9.5, 3.1) $\times$ 10$^{36}$ ergs~s$^{-1}$ at a distance
of 9.6~Mpc (P08).

Assuming that the narrow-line fluxes are powered by recombination, the
upper limit on the narrow H$\alpha$ flux sets a limit on the mass-loss
rate of the progenitor \citepeg{cd03}.  Using the measurement of the
shock velocity from the radio data of 52,000 km~s$^{-1}$ \citep{mv09}, 
we obtain a limit of $\dot{M} \lesssim$ 10$^{-5}$ M$_{\odot}$~yr$^{-1}$
($v_w$/10 km s$^{-1}$) (0.5/X$_H$), where $v_w$ is the velocity of
progenitor's wind and X$_H$ is the hydrogen mass fraction of the wind
material, which is likely to be lower than the solar value due to the
stripped nature of the progenitor.  This compares well with the
mass-loss rate directly estimated from the X-ray emission of \ax\ by
\citet{roming09} of $\dot{M} = (9\pm3) \times 10^{-6}$
M$_{\odot}$~yr$^{-1}$ ($v_w$/10 km s$^{-1}$), though they assumed that
the velocity of the SN shock was only 10,000 km s$^{-1}$.  This is
roughly an order of magnitude lower than the mass-loss rate of
SN~1993J \citep{roming09} and is consistent with the lack of other
signatures of CSM interaction in the spectra of \ax.

\subsection{Before Maximum Light}

\begin{figure}
\plotone{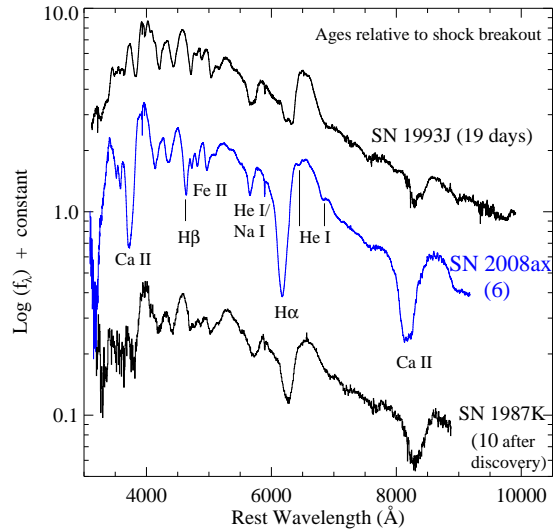}
\caption[Spectra of SNe~IIb in the hydrogen-dominated phase.]{Spectra
  of SNe~IIb in the hydrogen-dominated phase.  Important spectral
  features are labeled.  The SN~1993J spectrum is from 1993 Apr. 15
  \citep{fmh93} and a shock-breakout date of 1993 Mar. 27.5 has been
  assumed \citep{lewis94}.  The spectrum of SN~1987K is from 
  1987 Aug. 7 \citep{fil88} and the listed age is relative to the
  discovery on 1987 July 28 because the epoch of explosion is not
  tightly constrained.
}
\label{ax_IIbplot}
\end{figure}

The optical spectra of \ax\ evolved rapidly over the next 17 days on
the rise to maximum light, but were typical for a SN~IIb in the early
hydrogen-dominated phase (see also P08).  A comparison to the
prototypical SNe IIb 1987K \citep{fil88} and 1993J
\citep{nomoto93,schmidt93,fmh93,swartz93} at early times is 
shown in Figure~\ref{ax_IIbplot}.  The 
spectra continue to be dominated by strong Balmer absorption lines, as
well as the usual low-ionization species typical of SNe II (e.g.,
\ion{Ca}{2} and \ion{Fe}{2}).  \ion{Na}{1} D also likely contributes
to the absorption due to \ion{He}{1} $\lambda$5876.

In addition, notches have appeared in
the \hal\ profile due to \ion{He}{1} $\lambda$6678 and $\lambda$7065.
The day 2.34 spectrum did not show either line, but they developed
sometime between day 2 and day 5 (Fig.~\ref{ax_allplot}).  The day 2
spectrum did exhibit \ion{He}{1} $\lambda$5876
(Fig.~\ref{ax_earlyplot}), which was even stronger the previous night
\citep{blondin_cbet}.  The $\lambda$5876 line in the earliest spectra
of \ax\ was likely to be thermally excited, as is typical in very
young SNe II (cf. the other SNe~II in Fig.~\ref{ax_earlyplot}).
Thermal excitation would also explain why the strength of
$\lambda$5876 decreased from day 1 \citep{blondin_cbet} to day 2 as
the photospheric temperature cooled.  This effect is also seen in
normal SNe IIP, whose optical \ion{He}{1} lines disappear after the
photospheric temperature drops below $\sim$10,000~K
\citepeg{doug99em}.  The optical helium lines at 
later times in SNe Ib and IIb are excited by nonthermal electrons
created by exposure to $^{56}$Ni \citep{lucy91,swartz93}.

The appearance of \ion{He}{1} $\lambda$6678 and $\lambda$7065 in the
first week after explosion and their increasing strength thereafter
was a sign that the progenitor of \ax\ did not have the thick hydrogen
envelope of typical SNe~II and instead was partially stripped.
Supernovae which exhibit strong hydrogen lines at early times like
SNe~II but which transition to the helium-dominated spectra of SNe~Ib
at late times are known as SNe~IIb, a possibility that was first
predicted theoretically by \citet{woosley87}.  The most well-studied
example of an object undergoing such a transition was SN~1993J (see
Matheson et al. 2000a, and references therein).  \ax\ was more stripped
than SN~1993J and came to resemble a SN~Ib even sooner after
explosion (P08).

The premaximum NIR spectra of \ax\ (Fig.~\ref{ax_irplot}) are
remarkably smooth and devoid of prominent spectral features.  The rest
wavelengths of P$\alpha$ and P$\beta$ are marked with dotted lines and
question marks in Figure~\ref{ax_irplot}.  P$\alpha$ falls in a region
of strong telluric absorption between the $H$ and $K$ bands, but there
is not a prominent P-Cygni emission feature centered at that
wavelength.  At best, P$\beta$ contributes to a subtle wiggle in the
spectrum, but \ion{He}{1} 1.279~$\mu$m is close in wavelength and
likely contributes to that feature.  P08 attributed this feature
solely to P$\beta$ in their single NIR spectrum, taken on day 26 (7
days after $B$-band maximum light).  Br$\gamma$ at 2.166~$\mu$m also
does not appear to be present.  Collectively, evidence for hydrogen is
weak or lacking relative to the contemporaneous optical spectra.
Another piece of evidence against significant Paschen absorption is
that the features near P$\alpha$ and P$\beta$ did not weaken
substantially as the SN aged, unlike the optical Balmer lines.
Instead, the strongest features in the 1--2.5~$\mu$m spectral range
are absorptions from \ion{He}{1} 1.083 and 2.058 $\mu$m.  In addition
to those prominent features, the expected locations of absorptions
from other NIR \ion{He}{1} lines are marked in Figure~\ref{ax_irplot}
and they appear to match several of the weaker features.

The NIR spectra of SN~1993J were quite different.  While at the
earliest times the NIR spectra were very smooth and featureless
(similar to the optical spectra; Fig.~\ref{ax_earlyplot}), weak
P$\beta$ became apparent by about 4 days after shock breakout
\citep{wheel93,matthews02}.  P$\beta$ grew in prominence on the rise
to maximum light\footnote{In this work we use the term ``maximum
  light'' to refer to the \emph{second} peaks of the SNe~1993J and
  2005bf light curves.  At least in the case of SN~1993J, it is clear
  that the second peak is physically analogous to maximum
  light in the other objects \citepeg{nomoto93,pod93}.}
and beyond, peaking in strength 30--40 
days after explosion \citep{matthews02}.  The available NIR spectra of
SN 1993J are limited in wavelength coverage and spectral resolution,
so the appearance of the NIR helium lines is more difficult to
characterize, although \ion{He}{1} 1.083 and 2.058~$\mu$m were
definitely strong by 37 days after explosion
\citep{swartz93,matthews02}.  The lack of strong P$\beta$ in our
\ax\ spectra taken as soon as 10 days after shock breakout, and the
relative prominence of the helium lines being quite different from
SN~1993J at similar epochs, led \citet{marion08} to classify \ax\ as a
SN~Ib.

\subsection{Early Post-Maximum Phase}

\subsubsection{Spectral Comparisons}

\begin{figure*}
\plotone{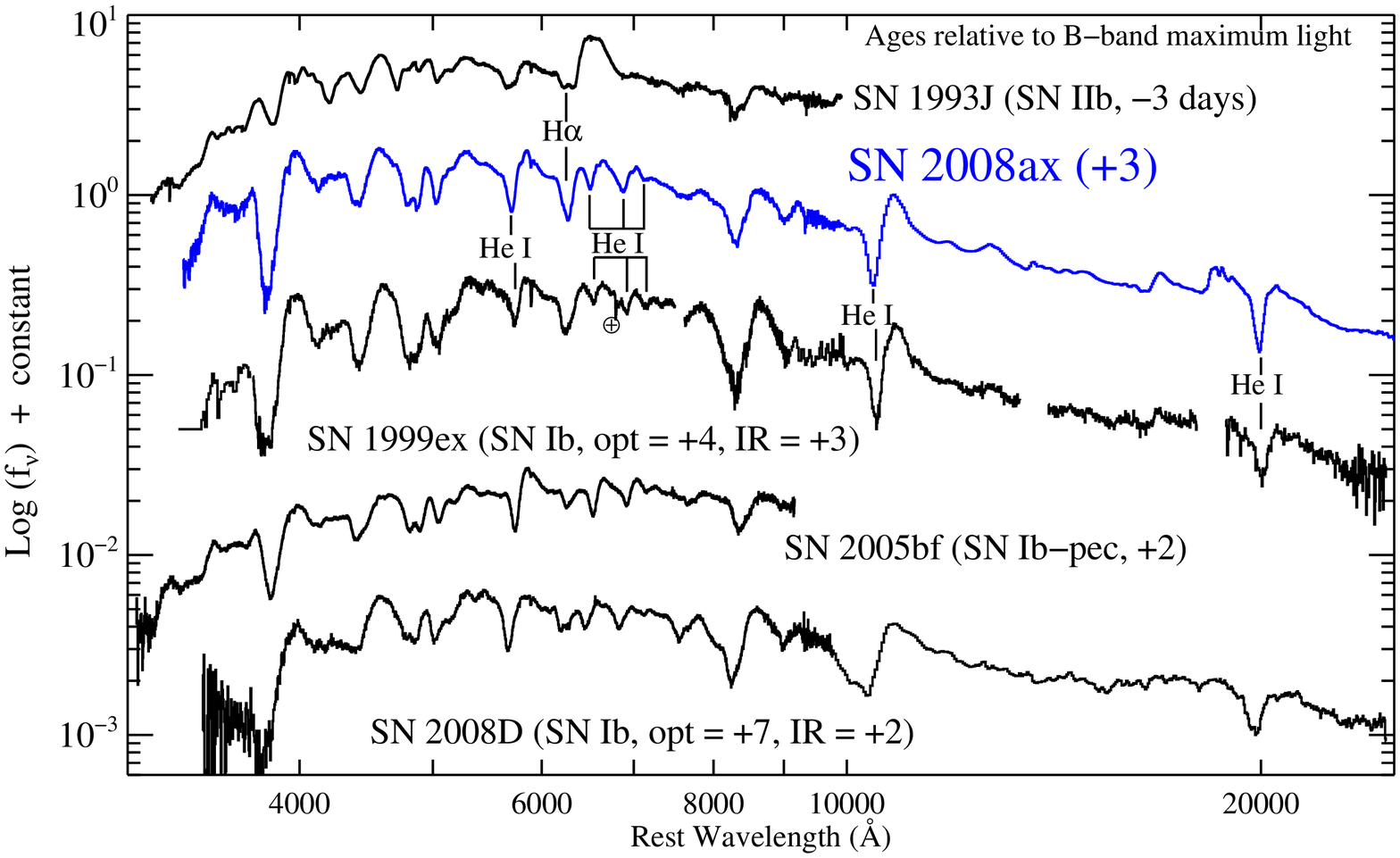}
\caption[Optical and NIR spectra taken near $B$-band maximum
  light.]{Optical and NIR spectra taken near $B$-band maximum
  light.  The ages in days relative to $B$-band maximum light are
  shown in 
  parentheses next to each spectrum.  The \ax\ spectra closely
  resemble those of the SNe~Ib 1999ex \citep{hamuy02}, 2005bf
  \citep{folatelli06}, and 2008D
  \citep{modjaz09}, while the SN~IIb 1993J  \citep{fmh93} shows
  significantly stronger H$\alpha$ and H$\beta$ lines.  If the 
  first spectrum of \ax\ had not been taken until maximum light, it
  likely 
  would have been classified as a SN~Ib like SN~1999ex.  The strongest
  \ion{He}{1} lines are marked, as is the H$\alpha$ line in SN~1993J
  and \ax.  Note that SN~1999ex likely also has H$\alpha$ absorption.
  A weak telluric absorption in the SN 1999ex spectrum is marked with
  a $\earth$ symbol.
}
\label{ax_maxplot}
\end{figure*}

The optical and NIR spectra of \ax\ near maximum light are
plotted in Figure~\ref{ax_maxplot} along with those of the SNe~Ib
1999ex \citep{hamuy02} and 2008D
\citep{soderberg08,modjaz09,mazzali08,malesani09}, as well as optical
spectra of the peculiar SN Ib/c 2005bf \citep{folatelli06} and the SN
IIb 1993J \citep{fmh93}.  SN~1993J stands out as 
the most distinct from the others because of its prominent
\hal\ emission. \ax\ and SN 1999ex are very similar over the full
optical through NIR spectral range, with the usual SN Ib P-Cygni
features of \ion{He}{1}, \ion{Fe}{2}, \ion{Ca}{2}, and \ion{O}{1} all
being present.  \ax\ also more closely resembles the SN Ib 2008D
than SN~1993J at this epoch.   

The largest difference between \ax\ and SN~2008D at this time is in
the absorption near 1.05~$\mu$m, which is clearly broader in
SN~2008D.  As discussed by \citet{taub06} and \citet{modjaz09},
SNe~Ib/c commonly exhibit an absorption near 1.04 $\mu$m that is
likely a blend of several species, with potential contributions from
\ion{Mg}{2}, \ion{Fe}{2}, \ion{C}{1}, \ion{Ca}{2}, \ion{S}{1}, and
\ion{Si}{1}, while only SNe Ib have strong absorptions from
\emph{both} of the \ion{He}{1} lines at rest wavelengths of 1.083 and
2.058~$\mu$m. The absorption near 1.05~$\mu$m in 
SN~2008D had a multi-component structure that was likely a blend
of both the usual 1.04~$\mu$m absorption blend and \ion{He}{1}
1.083~$\mu$m.  SNe~1999ex and 2008ax appear to have absorption from
only the \ion{He}{1} line, as the profile of the 1.05~$\mu$m
absorption is similar in velocity space (relative to 1.083~$\mu$m) to
that of the \ion{He}{1} 2.058~$\mu$m line, but with the flux
minimum being at a slightly higher velocity.  Both objects appear to
lack the additional component near 1.04~$\mu$m that results in the
broader appearance of the absorption in SN 2008D.

The similarity between the spectra of SNe 2008D and 2008ax near
maximum light is in contrast to their remarkably different spectra
shortly after explosion (Fig.~\ref{ax_earlyplot}).  Unfortunately, no
comparably early spectra of SN~1999ex exist.  That comparison would
have been especially interesting because the early-time light curve of
\ax\ during the envelope-cooling phase immediately following shock
breakout was more similar to that of SN~1999ex \citep{max02,roming09}.
The SN~Ib 2007Y also had an early-time light curve which closely
resembled SN~1999ex, and that object did show strong high-velocity
\hal\ absorption in the earliest spectrum, taken 14 days before
maximum light \citep{max09}. 
Despite the strength of the Balmer lines in the \ax\ spectra shortly
after explosion, the strongest evidence for hydrogen after maximum
light is the absorption due to \hal\ near 6270~\AA.

\subsubsection{\hal\ Absorption}

A robust detection of \hal\ absorption is an important insight from
our study of \ax.  Many SNe Ib have 
an absorption of variable depth near 6300~\AA\ for which several
identifications have been proposed.  For example, SN 1999ex clearly
shows such a feature in Figure~\ref{ax_maxplot}, similar to the one
seen in \ax.  The original study of \citet{hamuy02} identified it as
\ion{Si}{2} $\lambda6355$, but \citet{elmhamdi06} and
\citet{parrent07} argued that it was 
due to \hal.  \citet{branchIb} analyzed a large fraction of available
SN~Ib spectra and found that the 6300~\AA\ absorption was present in
most of the objects in their sample.  They examined several
possibilities for the identification of the line and found that
\hal\ was the most natural fit.  Our densely sampled spectral sequence
of \ax\ in Figure~\ref{ax_allplot} shows that the strong, obvious
\hal\ line at the earliest times smoothly evolved into the
6270~\AA\ absorption seen after maximum light before disappearing by
$\sim50$ days after explosion. 

\begin{figure}
\plotone{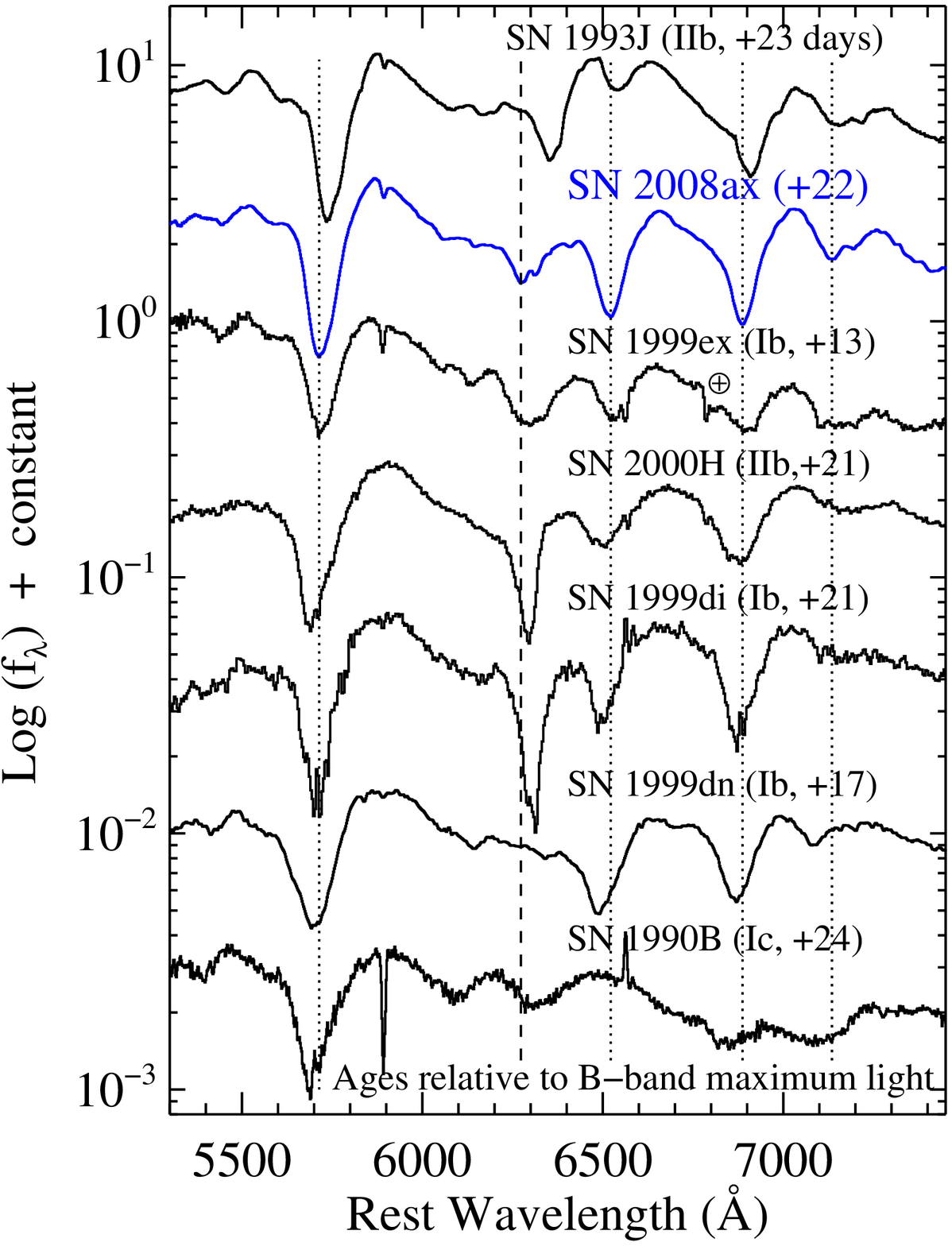}
\caption[Comparison of \ax\ to several stripped-envelope
  SNe after maximum light.]{Comparison of \ax\ to
  spectra of several stripped-envelope 
  SNe a few weeks after maximum light.  The vertical dashed line
  marks the location of H$\alpha$ in the spectrum of \ax\ and the
  vertical dotted lines mark the locations of the He I
  $\lambda\lambda$5876, 6678, 7065, and 7281 absorption minima in
  \ax.  The minima correspond to slightly different velocities
  depending on the optical depth in each line (see
  Table~\ref{ax_veltab}).  The comparison spectra come from two Type
  IIb supernovae (SN~1993J: Filippenko et al. 1994; SN~2000H: Branch
  et al. 2002), two ``hydrogen-rich'' SNe~Ib
  (SN 1999di: Matheson et al. 2001; SN~1999ex: Hamuy et al. 2002), one
  normal SN~Ib (SN~1999dn, which did have \hal\ absorption at earlier
  times: Matheson et al. 2001), and a normal SN~Ic (SN~1990B:
  Clocchiatti et al. 2001).
  A telluric absorption in the SN 1999ex spectrum is marked with a
  $\earth$ symbol.
}
\label{ax_heiplot}
\end{figure}

We have plotted a comparison of the post-maximum-light spectra of
several of the hydrogen-rich SNe~Ib and SNe IIb from \citet{branchIb}
in Figure~\ref{ax_heiplot}.  The \hal\ absorption of \ax\ is not
notably strong relative to the other objects, in contrast with the
strength of Balmer lines in \ax\ relative to other core-collapse SNe
in the earliest spectra (Fig.~\ref{ax_earlyplot}).  Most SNe are not 
observed as soon after explosion as \ax.  In many cases, the first
spectrum of a supernova is taken around or shortly after maximum
light.  If the first spectrum of \ax\ had not been taken until after
maximum light, it probably would have been classified as a SN~Ib like
SN 1999ex.  Similarly, the time evolution of the radio emission from CSM
interaction in \ax\ is more typical of SNe~Ib/c than the strong radio
emission seen in SN~1993J \citep{roming09}.

We used the SuperNova IDentification program (SNID; Blondin \& Tonry
2007) to cross-correlate the near-maximum-light (day 22) optical
spectrum of \ax\ with the default library of supernova templates that
are distributed with SNID.  The best
matches were all with SNe 2005bf \citep{tominaga05,folatelli06} and
1999ex \citep{hamuy02} near maximum light.  SNID also finds matches
for the day 14 and 27 spectra with SNe IIb such as SN 1996cb
\citep{qiu99} and SN 2000H \citep{branchIb} or the hydrogen-rich SNe
Ib 1999dn and 1999di \citep{ma01}.  By day 41 (22 days after $B$
maximum), the best SNID matches are many of the same SNe Ib as shown
in Figure~\ref{ax_heiplot}.  These SNID-based comparisons (using the
default set of templates) are probably representative of the likely
spectroscopic classification
of \ax\ in the International Astronomical Union Circulars had it not
been discovered so soon after explosion.  

This suggests that a substantial fraction of objects classified as SNe
Ib might have similar amounts of hydrogen in their outer ejecta as
\ax.  Another example is the SN~Ib 2007Y \citep{max09}, whose earliest
spectra show strong high-velocity \hal\ absorption, but by maximum
light also closely resembled SN~2005bf.  An analysis of the nebular
spectra by \citet{mau10} concluded that SN~2007Y and \ax\ probably
ejected similar amounts of hydrogen.  As discussed by
\citet{branchIb}, this has important implications for the nature of
their progenitors.  Additionally, the division between spectroscopic
classifications of objects as SNe Ib vs. IIb may in some cases depend
on the timing of the initial spectroscopy, although not for those SNe
whose progenitors retain as much hydrogen as SN 1993J.

While there is a small possibility that some other line is blended
with \hal\ and comes to dominate the absorption feature at late times, 
we believe the simplest explanation is that it is due to \hal\ at all
times.  As an example, \citet{deng00} claimed that the increasing
prominence of \ion{C}{2} $\lambda$6580 relative to \hal\ explained the
time evolution of the 6300~\AA\ feature in the SN Ib 1999dn, but
\citet{branchIb} presented arguments against the presence of strong
\ion{C}{2} in that particular object, including the weakness of any
\ion{C}{2} $\lambda$4745 absorption.  In addition, the lack of
\ion{C}{1} lines in the NIR spectra of \ax\ (see below) suggests that
carbon likely does not make a strong contribution to the optical
spectrum as well. 

\subsubsection{Velocity Evolution}

\begin{figure}
\plotone{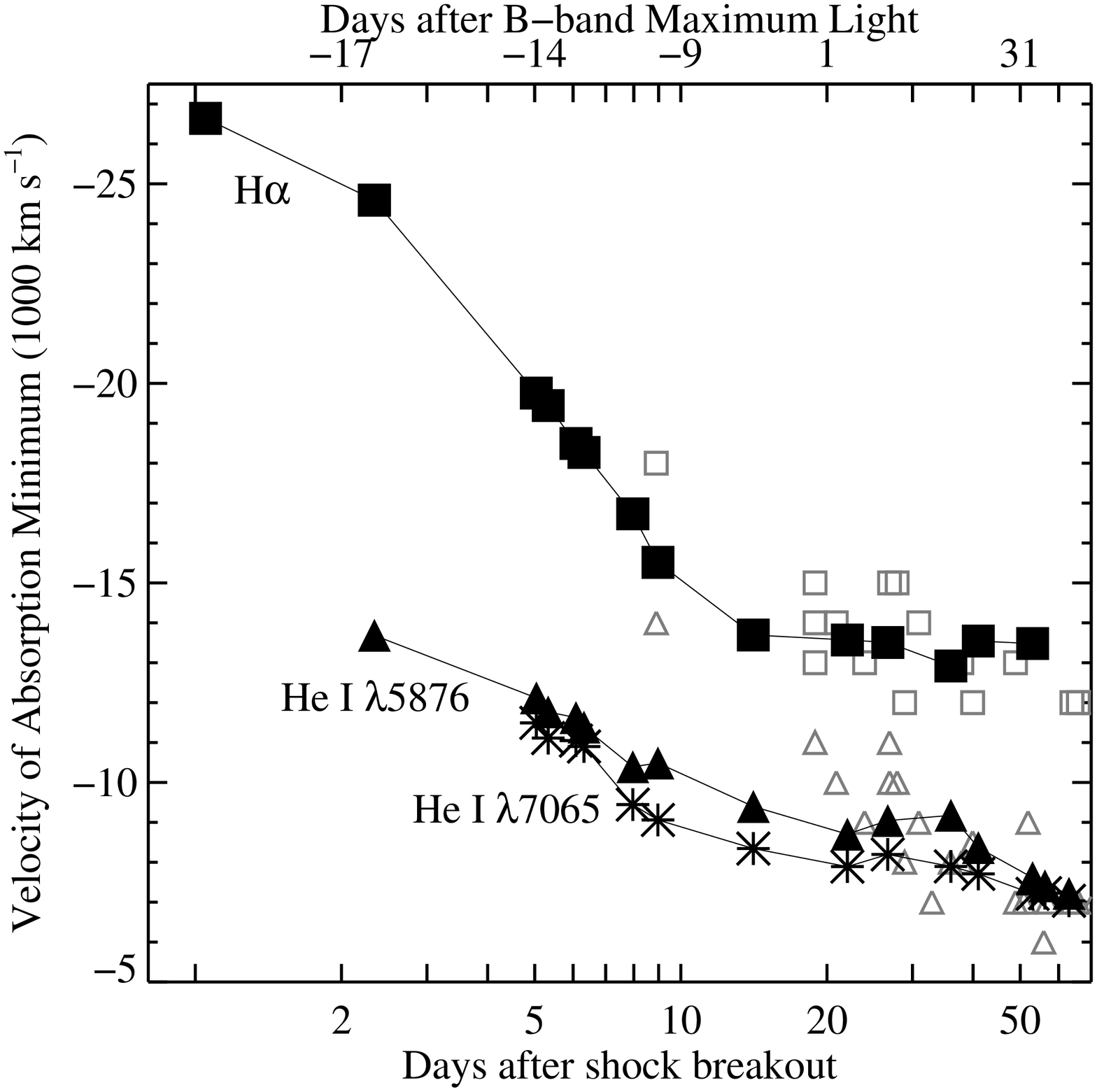}
\caption[Time evolution of the velocities of absorption minima in
  \ax.]{Velocities of the absorption minima of H$\alpha$ (\emph{filled 
  squares}), \ion{He}{1} $\lambda$5876 (\emph{filled triangles}), and
  \ion{He}{1} $\lambda$7065 (\emph{black stars}) for \ax.  The first
  \hal\ data point (at day 1.05) is from \citet{blondin_cbet}.  Other
  data points in gray represent 
  velocities for hydrogen (\emph{open squares}) and \ion{He}{1}
  (\emph{open triangles}) 
  for various SNe~Ib taken from the \synow\ fits of \cite{branchIb}.
  The points from \cite{branchIb} are listed relative to maximum light
  and have been shifted assuming the SNe all had the same risetime
  as \ax\ of 19 days.  The velocities for \ax\ are consistent with
  those of the other objects in the literature.
}
\label{ax_velplot}
\end{figure}

Further evidence to support the \hal\ identification for the
6270~\AA\ feature in \ax\ can be
seen from its velocity evolution, which we have listed
in Table~\ref{ax_veltab} and plotted in Figure~\ref{ax_velplot}.  Our
velocity measurements for the absorption minima are derived by taking
the wavelength of the minimum flux of a spline fit to the absorption
profile and applying the relativistic Doppler formula for the
velocity. \citet{branchIb} used their \synow\ code to fit the spectra 
of a sample of SNe~Ib and derived minimum velocities for \hal\ and
\ion{He}{1}, which are plotted in Figure~\ref{ax_velplot} as open gray 
boxes and triangles, respectively, assuming that the 6270~\AA\ feature
is always due to \hal.  In order to plot the literature points on the
same graph as \ax\ when we do not know the dates of explosion of the
other objects, we have assumed a common rise time to maximum
light, which is likely incorrect at the level of a few days offset per
object due to the dispersion in SN~Ib rise times.  The
post-maximum-light \hal\ and helium velocities in \ax\ are consistent
with the normal SNe~Ib. 

The very high early-time velocity of the \hal\ absorption minimum in
\ax\ rapidly declined until day 14, when it started to asymptotically
approach a value of $-$13,500 \kms, which is well above the
photospheric and helium velocities, but within the range of variation
shown by the other objects.  An identification of the
6270~\AA\ feature with either \ion{C}{2} $\lambda$6580
\citepeg{deng00} or \ion{Si}{2} $\lambda$6355 \citepeg{hamuy02} would
have to explain the unusual asymptotic velocities ($-$14,300 and
$-$3900 \kms, respectively) of those elements.  Even if \hal\ were
blended with absorptions from those ions, the late-time velocities are
a strong argument that \hal\ must be dominant.

We have not plotted the \hbeta\ velocities in Figure~\ref{ax_velplot}
because \hbeta\ was completely dominated by \ion{Fe}{2} lines after
about day 9, but our last solid detection on day 9 also had an
absorption minimum at about $-13,000$ \kms.  The most natural
explanation is that the 6270~\AA\ feature is indeed due to \hal\ and
that it formed in a low-mass hydrogen-rich layer in the outer ejecta.
The hydrogen velocities never dropped below $-$13,000 \kms\ because
there was no hydrogen interior to that in the ejecta. 

\citet{branchIb} also point out
that lines forming at significantly greater velocities than the
continuum photosphere (a geometry they term ``detached'' line
formation) will have a characteristic profile with deep P-Cygni
absorption and a weak emission component.  This seems consistent with
the unusually deep absorption components of the Balmer lines in
\ax\ at early times (Fig.~\ref{ax_earlyplot}) and the lack of an
emission component to \hal\ at later times.

\begin{deluxetable*}{llccccc}
\tabletypesize{\scriptsize}
\tablecaption{Velocities of Absorption Minima}
\tablehead{\colhead{$\Delta$t\tablenotemark{a}} & 
\colhead{$t_{B_{\rm{max}}}$\tablenotemark{b}} &
\colhead{\hal} & \colhead{He I $\lambda$5876} & 
 \colhead{He I $\lambda$7065} &
 \colhead{He I 1.0830~$\mu$m} &
 \colhead{He I 2.0581~$\mu$m} \\
(days) & (days) & (\kms) & (\kms) & (\kms) & (\kms)& (\kms)}
\startdata
2.34 & -16.5 & $-$24,600 & $-$13,700 & \nodata & \nodata &\nodata \\
5.04 & -13.8 & $-$19,800 & $-$12,100 & $-$11,500 & \nodata &\nodata \\
5.33 & -13.6 & $-$19,400 & $-$11,800 & $-$11,100 & \nodata &\nodata \\
6.12 & -12.8 & $-$18,500 & $-$11,600 & $-$11,000 & \nodata &\nodata \\
6.32 & -12.6 & $-$18,300 & $-$11,400 & $-$10,900 & \nodata &\nodata \\
7.96 & -10.9 & $-$16,700 & $-$10,400 & $-$9,400 & \nodata &\nodata \\
8.97 & -9.9 & $-$15,500 & $-$10,500 & $-$9,100 & \nodata &\nodata \\
10.2 & -8.7 & \nodata & \nodata & \nodata & $-$11,700 & $-$10,000 \\
11.2 & -7.7 & \nodata & \nodata & \nodata & $-$11,300 & $-$9,800 \\
14.1 & -4.8 & $-$13,700 & $-$9,400 & $-$8,300 & \nodata &\nodata \\
22.0 & 3.1 & $-$13,600 & $-$8,700 & $-$7,900 & \nodata &\nodata \\
22.1 & 3.2& \nodata & \nodata & \nodata & $-$10,700 & $-$9,300 \\
27.2 & 8.3 & $-$13,500 & $-$9,000 & $-$8,200 & \nodata &\nodata \\
36.0 & 17.1 & $-$12,900 & $-$9,200 & $-$7,900 & \nodata &\nodata \\
40.2 & 21.3 & \nodata & \nodata & \nodata & $-$11,900 & $-$9,300 \\
41.0 & 22.1 & $-$13,500 & $-$8,400 & $-$7,700 & \nodata &\nodata \\
53.0 & 34.1 & $-$13,500 & $-$7,600 & $-$7,200 & \nodata &\nodata \\
56.2 & 37.3 & \nodata & $-$7,400 & $-$7,200 & \nodata &\nodata \\
63.0 & 44.1 & \nodata & $-$7,200 & $-$7,000 & \nodata &\nodata \\
66.0 & 47.1 & \nodata & \nodata & \nodata & $-$11,100 & $-$9,300 \\
\enddata
\tablenotetext{a}{Time after shock breakout.}
\tablenotetext{b}{Time after maximum light in the $B$ band.}
\label{ax_veltab}
\end{deluxetable*}

\subsubsection{NIR Spectra}

Next we consider the post-maximum-light evolution of our NIR
spectra.  The early-time spectra were relatively smooth with most 
features attributable to \ion{He}{1} lines, but
after maximum light many more features appeared, most of which are
rather subtle blends of emission from neutral or singly ionized
intermediate-mass elements such as sodium, silicon, magnesium, and
calcium \citepeg{meikle89,gerardy04b}.  The most 
prominent of these is an emission feature from \ion{Mg}{1} at
1.503~$\mu$m.  The strongest expected NIR lines from iron-peak
elements are [\ion{Fe}{2}] 1.257 and 1.644~$\mu$m \citep{meikle89}.
Our spectra do not show those features, but this may not be
significant because our last epoch of observation (day 66) is before
the full transition to the nebular phase and subsequent emergence of
the forbidden lines.  We defer detailed analysis of the relative
contributions of each element to the blends to future work on the
analysis of the NIR spectra of SNe Ib/c (Marion et al., in prep.).

We do wish to examine one issue, however, which is the relative
strength of carbon and oxygen lines.  \ion{O}{1} lines, in particular
$\lambda$7774, $\lambda$9266, and 1.129~$\mu$m, are present at
late times in \ax\ (Fig.~\ref{ax_irplot}).  The strong \ion{Ca}{2} NIR
triplet absorption prevents us from determining the presence of any
contribution from \ion{O}{1} $\lambda$8446.  \ion{C}{1} lines have been
strongly detected in some SNe Ic, such as SN~1994I \citep{sauer06} and
SN~2007gr \citep{valenti08,hunter09}.  However, most stripped-envelope
SNe do not show such lines, including other SNe~Ic such as SN
2002ap \citep{gerardy04b}. The strong \ion{C}{1} lines near $\lambda$9094,
$\lambda$9406, $\lambda$9658, and 1.069~$\mu$m fall in the overlap
region between our optical and NIR spectra of \ax\ and clearly are not
present in either. 
\citet{gerardy04b} identified an emission feature from \ion{C}{1} at
1.454~$\mu$m at 30 days after maximum light in the SN~Ic 2000ew that
was almost as strong as the neighboring \ion{Mg}{1} 1.503~$\mu$m
emission line, but no such feature is visible in our \ax\ NIR 
spectra at comparable epochs (Fig.~\ref{ax_irplot}).  The reason why
some SNe~Ic have significantly stronger carbon lines
than other stripped-envelope SNe remains unclear
\citep{valenti08,hunter09}.

\subsection{Late-Time Spectra}

\begin{figure}
\plotone{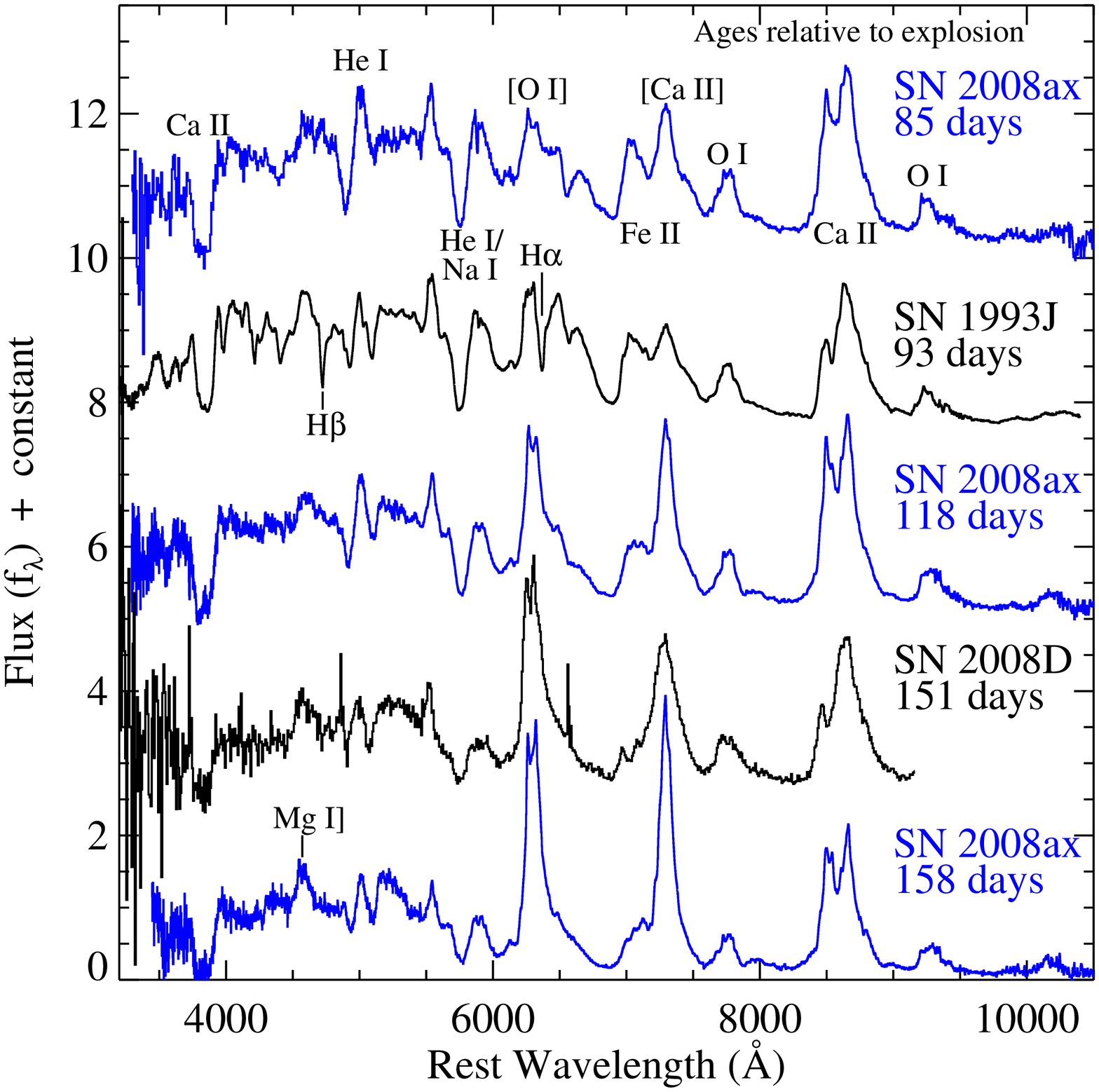}
\caption[Late-time spectra of \ax.]{Late-time spectra of \ax\ during
  the transition to the nebular phase.  Prominent spectral features are
  labeled.  The spectrum of SN~1993J is from 1993 June 28
  \citep{fmb94} and still shows evidence of hydrogen, with the Balmer
  absorption lines labeled.  No strong hydrogen lines are evident in the
  \ax\ spectra, nor in the SN~2008D spectrum from 2008 June 7
  \citep{modjaz09}.  An additional total-flux spectrum of \ax\ was
  obtained on day 99 in spectropolarimetry mode and is discussed in
  \S4. 
}
\label{ax_lateplot}
\end{figure}

After maximum light, the rapid spectral evolution seen at earlier
times slowed down.  P-Cygni features from \ion{He}{1} lines (4471,
5015, 5876, 6678, 7065, and 7281 \AA) all became prominent in the flux
spectrum  (Fig.~\ref{ax_allplot}), and the emission component of the
\ion{Ca}{2} NIR triplet began to dominate the line profile.  The
absorption component of the NIR triplet largely disappeared between
days 41 and 53, while the \hal\ absorption near 6270~\AA\ became hard
to discern at the same time.  These spectral changes coincide with the
light curve of \ax\ transitioning from the main peak to the late-time
radioactive-decay tail (P08; Roming et al. 2009).  The late-time
spectra of \ax, starting 85 days after explosion, are plotted in
Figure~\ref{ax_lateplot}.  Over the next 70 days, the spectra make the
transition from the photospheric to the nebular phase.  The P-Cygni
absorption features weaken and strong emission lines from oxygen and
calcium start to dominate the spectrum.

For comparison, a spectrum of SN~1993J at similar epochs is also
plotted.  Balmer lines are still present in the SN~1993J spectrum.
\hal\ never disappeared from the SN~1993J spectra, even as \hal\ faded
in the first several months after maximum light \citep{fmb94}.  \hal\
subsequently rose in strength in SN~1993J as the interaction with
hydrogen-rich CSM began, but our \ax\ spectra do not extend to such
late times.

Any \hal\ emission at late times must be hidden in the far red wing of
the [\ion{O}{1}] $\lambda\lambda$6300, 6364 line profile.
\citet{mili09} do identify \hal\ in that red wing and there does
appear to be emission with a box-like profile present in their
day 307 spectrum, similar to SN~1993J at late times, albeit
at a greatly reduced level.  However, any \hal\ present at the epoch
of our data must be blended with other lines, as there is little
difference in the red wing of the [\ion{O}{1}] line profile between
\ax\ and the SN~2008D spectrum from a similar epoch \citep{modjaz09}.
The biggest difference between those two objects at these late times
is that SN 2008D shows broader emission lines.  Electron scattering
can also naturally explain red wings of the nebular line profiles
\citep{fc89}.  In short, we do not find any compelling evidence for
\hal\ emission in our spectra of \ax\ at late times.  However, spectra
taken at even later times do appear to show the emergence of box-like
\hal\ emission indicative of CSM interaction
\citep{mili09,mau10,taub11}.

\begin{figure}
\plotone{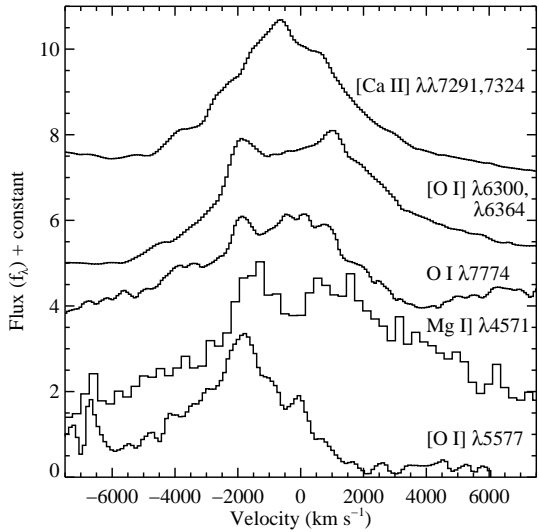}
\caption[Late-time emission-line profiles.]{Late-time emission-line
  profiles of 
  \ax\ on day 158, our latest epoch of observations.  The
          [\ion{Ca}{2}] line profile (presented relative to the
          average wavelength of the doublet of 7308~\AA) is largely
          symmetrical.  The [\ion{O}{1}] $\lambda\lambda$6300, 6364
          emission (presented relative to $\lambda$6300) is
          double peaked, but other oxygen lines do not 
          share the same line profile.
}
\label{ax_oiplot}
\end{figure}

We have plotted the profiles of strong emission lines in our
latest \ax\ spectrum, taken on day 158, in Figure~\ref{ax_oiplot}.
The oxygen lines collectively have rather different line profiles from
each other and from the [\ion{Ca}{2}] $\lambda\lambda$7291, 7324
emission.  Interest in the late-time line profiles of
stripped-envelope SNe has risen in recent years after \citet{maz03jd} 
identified double-peaked oxygen and magnesium lines in SN~2003jd and
interpreted them as being due to an off-axis jet-like explosion.
Subsequent work \citep{maeda08,modjaz08,taub09,mili09,maurer09} has
shown that such profiles are ubiquitous, although the interpretation
is uncertain.  In particular, \citet{mili09} have presented an
extensive study of the late time emission-line profiles of \ax\ along
with comparisons to several other stripped-envelope SNe. 

In the case of \ax, the [\ion{O}{1}] $\lambda\lambda$6300, 6364 line
profile is clearly double peaked, with the peaks separated by
$\sim$3000 \kms.  The \ion{O}{1} $\lambda$7774 emission exhibits a
flatter top, but with a definite emission peak under the bluer of the
two peaks seen in [\ion{O}{1}] $\lambda\lambda$6300, 6364 (some of the
weaker wiggles seen in the \ion{O}{1} $\lambda$7774 profile are likely
due to poor flatfielding).  The [\ion{O}{1}] $\lambda$5577 emission
line has a single blueshifted peak, again at a 
radial velocity of $-2000$~\kms.  Although our data at bluer
wavelengths are very noisy, the \ion{Mg}{1}] $\lambda$4571 emission
feature does appear to be double peaked as well.  This represents a
difference with the spectra presented by \citet{mili09}. 
Their spectra showed a single blueshifted peak for the \ion{Mg}{1}]
  $\lambda$4571 line, similar to the [\ion{O}{1}] $\lambda$5577
  emission  at early times.  However, their earliest spectrum
  including $\lambda$4571 was taken on day 307,
potentially indicating that the line profile underwent significant
evolution. 

\citet{mili09} found that toroidal models for the oxygen-emitting zone
had problems explaining the line profiles of most stripped-envelope
SNe.  They explored two alternatives in the case of \ax.  The first
invoked a blueshifted oxygen distribution that was optically thick so
that $\lambda$6300 and $\lambda$6364 had a 1:1 ratio.  The second
had a largely symmetric broad oxygen distribution with an additional
blueshifted clump producing the peaks in the profile.  Both models
still have some drawbacks, leaving the geometry highly uncertain.  Our
\ion{Mg}{1}] $\lambda$4571 line profile at the early nebular phase
indicates that the magnesium distribution (which is thought to trace
the oxygen distribution as well; Taubenberger et al. 2009) is not
highly blueshifted, disfavoring the 
first model of \citet{mili09}.  Regardless of the actual geometry,
these asymmetric and double-peaked oxygen line profiles are a sign of
strong asphericities in the core of \ax.

\section{Spectropolarimetry\label{ax_specpol}}

Thomson scattering in hot supernova atmospheres polarizes outgoing
radiation.  Spatially unresolved SNe will show no net polarization if
they are circularly symmetric when projected on the plane of the sky.
However, if they are aspherical the net continuum polarization will
be nonzero \citep{ss82}.  Line opacity is generally thought to be
depolarizing, so polarization modulations across spectral features can
trace the aspherical distribution or excitation of individual chemical
elements within the ejecta.  The expectation is that line scattering
in an axisymmetric, but aspherical, electron-scattering SN atmosphere
will produce ``inverted'' P-Cygni polarization modulations
\citepeg{je91}.  Unpolarized line emission near zero velocity dilutes
the continuum polarization down to almost zero while selective
blocking of forward-scattered continuum light results in a
polarization maximum associated with the flux minimum.

Detailed analysis of the Stokes parameters $q$ and $u$ can also test
whether any asphericities present are axisymmetric.  In axisymmetry,
data points representing bins in wavelength will lie along a 
line in the $q$-$u$ plane, as they represent greater or lesser amounts
of polarization with a consistent orientation.  A more complex
distribution of wavelength bins in the $q$-$u$ plane is indicative of
non-axisymmetric asphericities.  In particular, if the wavelength bins
associated with spectral lines from different ions have distinct
behaviors, then the two ionic species likely have different spatial
distributions within the ejecta.  Thus, we will be able to compare the
structure of the ejecta of \ax\ with those of previously studied
stripped-envelope SNe \citep{ww08}.  While we can examine the relative
distributions of different chemical elements and make comparisons to
other objects, spectropolarimetric data cannot be inverted to directly
determine the actual geometry.

Core-collapse SNe occurring in progenitor stars that have
retained most of their massive hydrogen envelopes explode as normal
SNe II.  Well-studied examples, such as the SN IIP 1999em
\citep{le01}, show only small values of the polarization at early
times while the photosphere recedes through the massive and apparently
almost spherical hydrogen envelope.  However, evidence is mounting
that even those core-collapse explosions have highly aspherical cores
\citep{doug04dj,meIIP}.  \citet{wa01} identified a 
trend where the observed amount of polarization in core-collapse
SNe increases both with time, as the centers of the explosions
become visible, and with increasing amounts of stripping of the
progenitor stars.  Therefore, it should not be a surprise that a SN
whose progenitor was as highly stripped as that of \ax\ should show a
strong polarization signal even at early times.

\begin{figure*}
\plottwo{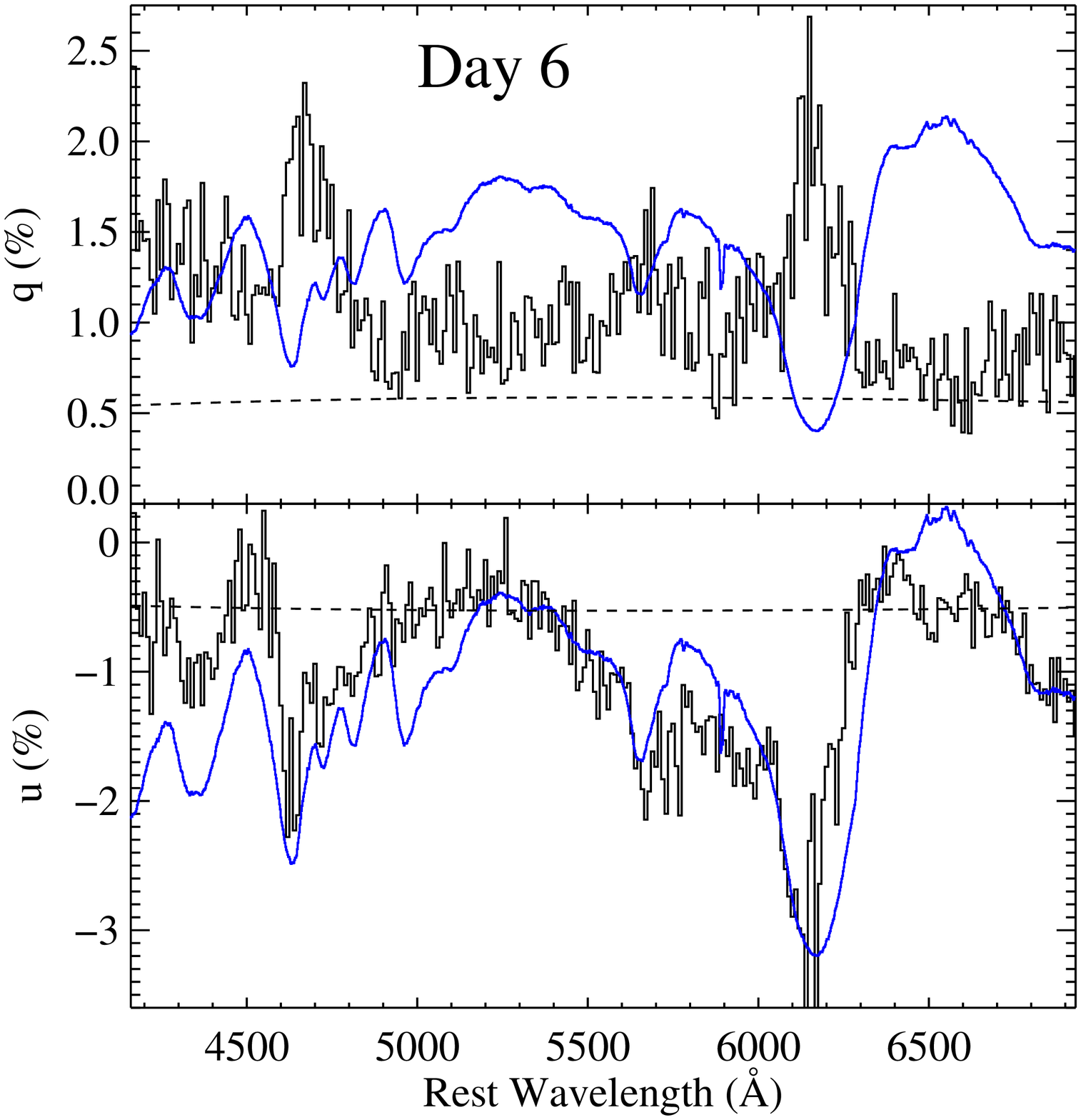}{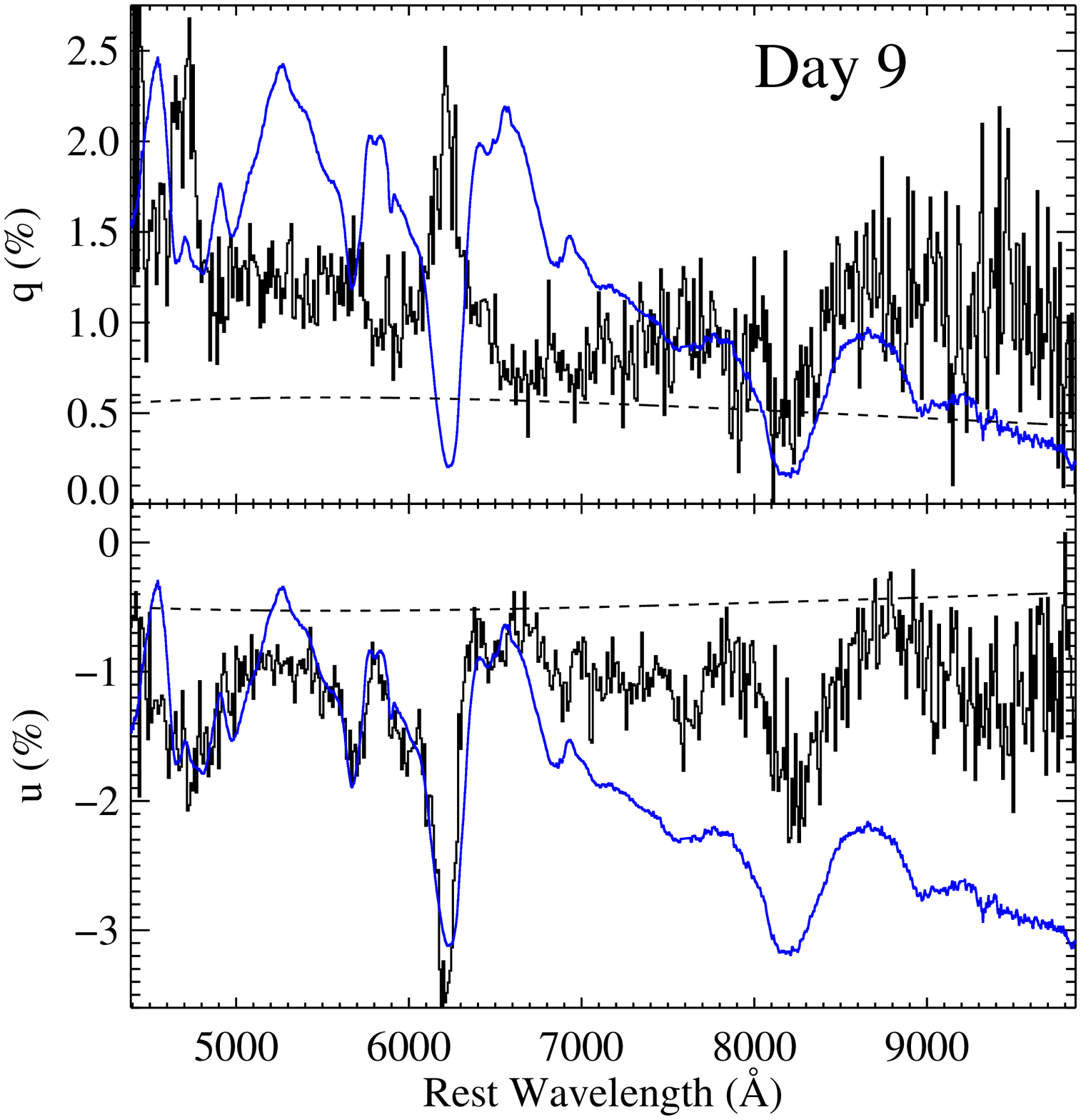}
\caption[Early-time spectropolarimetry of \ax.]{Early-time
  spectropolarimetry of \ax, uncorrected for ISP, and plotted in
  10~\AA\ bins.  Day 6 is on the left and day 9 is on the right.  The
  gray line in each panel (blue in the online edition) is the total-flux
  spectrum (in $f_{\lambda}$, with arbitrary multiplicative and
  additive offsets applied) from each night to guide the eye.  The
  dashed lines show our preferred ISP as a reference.  Strong
  polarization modulations are associated with the line features in
  the total-flux spectrum.
}
\label{ax_rawpolplot}
\end{figure*}

We have plotted $q$ and $u$ from our two epochs
of early-time (days 6 and 9) spectropolarimetry of \ax\ in
Figure~\ref{ax_rawpolplot}.  High levels of polarization are present,
with observed polarizations in the rest-frame $V$ band of $1.37 \pm
0.04$\% and $1.61 \pm 0.04$\% on days 6 and 9, respectively.  This
change in polarization in only three days, along with the very large
polarization modulations across spectral features such as \hal, is
evidence of substantial intrinsic polarization in \ax.

\subsection{Interstellar Polarization Correction}

We should expect that interstellar polarization (ISP) from dust in the
host galaxy of \ax\ makes a significant contribution to the 
observed polarization, so we wish to remove this effect before
studying its intrinsic polarization properties.  Typical
values for the polarization efficiency of Galactic dust \citep{ser75}
imply a potential host ISP of 1.5\% for our adopted reddening.  The
Galactic component of the reddening should contribute negligibly to
the ISP.  In order to separate intrinsic SN polarization from ISP, we
must make some assumptions.  Early work on 
the SN IIb 1993J assumed that the intrinsic polarization in the
emission component of \hal\ should be zero \citep{thw93,ho96,tran97}.
\citet{maund01ig} showed that this assumption was false, at least in
the case of the peculiar SN IIb 2001ig, in part due to blending of
\ion{He}{1} $\lambda$6678 with the peak of \hal.  Our spectra of
\ax\ at these early times (gray lines in Fig.~\ref{ax_rawpolplot},
blue in the online version) do
show developing \ion{He}{1} lines, so we will heed the cautionary 
example of SN 2001ig and use another method to estimate the ISP.  

\citet{maund01ig} were able to robustly measure the ISP to SN
2001ig from spectropolarimetry obtained on day 256.  At that late
epoch, the optical depth to electron scattering in the SN ejecta
should be small as the expansion of the SN and consequent decrease in
electron density results in geometric dilution of the scattering
probability.  As a consistency check, no polarization modulations were
seen across SN spectral features, indicating that the measured
polarization was entirely due to ISP.  A similar method was
successfully used by \citet{je91catalog} to correct the collected SN
1987A polarization dataset for ISP.

\begin{figure}
\plotone{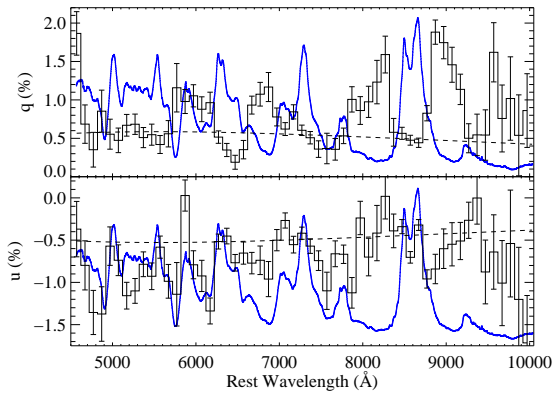}
\caption[Polarization of \ax\ on day 99.]{Polarization
  of \ax\ on day 99 after explosion, uncorrected for ISP.  The two
  panels show the observed Stokes parameters $q$ (top) and $u$
  (bottom) plotted versus wavelength compared to the total-flux
  spectrum of \ax\ (gray lines, blue in the online version).
  Depressions in polarization are visible at the wavelengths 
  of the strong nebular emission lines, such as the \ion{Ca}{2} NIR
  triplet near 8600 \AA.  The dashed lines are our preferred ISP,
  chosen to make the intrinsic polarization nearly zero at the peaks
  of the 
  nebular emission lines.  These variations in polarization across
  spectral features mean that \ax\ must have intrinsic continuum
  polarization in excess of 1\% even at this late date, and hence large
  central asphericities. 
}
\label{ax_latepolplot}
\end{figure}

Our final epoch of spectropolarimetry of \ax\ was taken on day 99 and
is plotted in Figure~\ref{ax_latepolplot}.  Strong polarization
modulations are still seen at this late time, particularly across the
\ion{Ca}{2} NIR triplet emission feature near 8600~\AA, indicating
that there are still both substantial asphericities present in the
ejecta and a sufficient optical depth to electron scattering to add a
component of intrinsic SN polarization to the ISP to form the
polarization signal we measure.  The
pattern of the polarization modulations, with decreases in
polarization present at the strong emission lines (particularly in
$q$), suggests a method to decompose the polarization.

We expect that the continuum polarization we see at this late time is
due to photospheric continuum light scattering off electrons
relatively deep in the ejecta.  The strong emission lines which become
more prominent as the supernova transitions to the nebular phase are
mostly due to material located exterior to the region of highest
optical depth to electron scattering.  The intrinsically unpolarized
photons in the lines will travel unimpeded to the outside observer,
acquiring a polarization only from the ISP along the line of sight.  A
similar dilution effect was seen by \citet{meIIP} in
the emission components of nebular lines such as [\ion{Ca}{2}] in
SNe~IIP at late times. 

With that in mind, we selected three wavelength regions of
6230$-$6350, 7250$-$7370, and 8470$-$8730~\AA\ to represent the peaks
of strong emission features from (respectively) [\ion{O}{1}] $\lambda$6300,
[\ion{Ca}{2}] $\lambda$7300, and the \ion{Ca}{2} NIR triplet.
Integrating the observed polarization over those three windows gave
consistent values for the Stokes parameters, with a spread of
$\sim$0.1\% in each, indicating that the peaks of the three lines do
have similar polarizations of about 0.7\%.  We fit an ISP curve to
20~\AA\ bins of the Stokes parameters from those three regions,
following the functional form of \citet{ser75} with the modifications
by \citet{whittet92}.  We fixed the peak wavelength of the ISP curve
to 5550~\AA\ (or, equivalently, we set $R_V = 3.1$) because the limited
number of data points being fit were all at longer wavelengths than
the expected peak, so the peak wavelength was not well constrained.
Our measured values for ($q_{\rm{ISP}}$, $u_{\rm{ISP}}$) are (0.59\%,
$-$0.52\%), which we adopt whenever we correct our data for ISP below,
along with the functional form for ISP of \citet{whittet92}.  This ISP
value has an uncertainty of 0.1\% in each Stokes parameter based on
the spread in the polarizations at the three line peaks, which is
probably more realistic than the small formal errors from the fit
\citep{maund01ig}.

\begin{figure}
\plotone{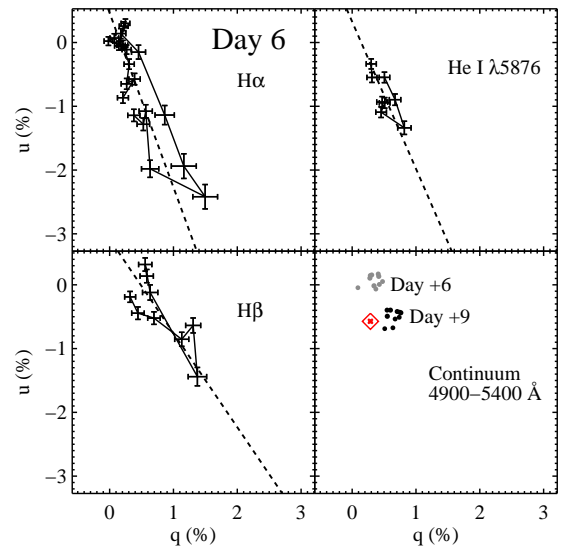}
\caption[Day 6 line-polarization data plotted in the $q-u$
  plane]{Line-polarization data from day 6 plotted in the $q-u$ plane,
  after subtraction of ISP.  The data points with error bars in black
  in each panel represent 50~\AA\ bins in wavelength ranges
  corresponding to the indicated spectral features (H$\alpha$:
  5900$-$6950~\AA; H$\beta$: 4450$-$4900~\AA; \ion{He}{1}:
  5450$-$5850~\AA).  The solid lines connect the points in order of
  wavelength to better show the loop structure.  The overplotted
  dashed lines represent 
  linear fits to the line-polarization data to guide the eye.
  The lower-right panel shows a comparison of data points
  taken from the 4900$-$5400 \AA\ region on both nights,
  exhibiting a net shift of 0.64\%.  Also plotted in that panel is the
  7000$-$7450~\AA\ continuum polarization from day 9 (diamond, red in
  the online version) for comparison with Figure~\ref{ax_n2lineplot}.
}
\label{ax_n1lineplot}
\end{figure}

\begin{figure}
\plotone{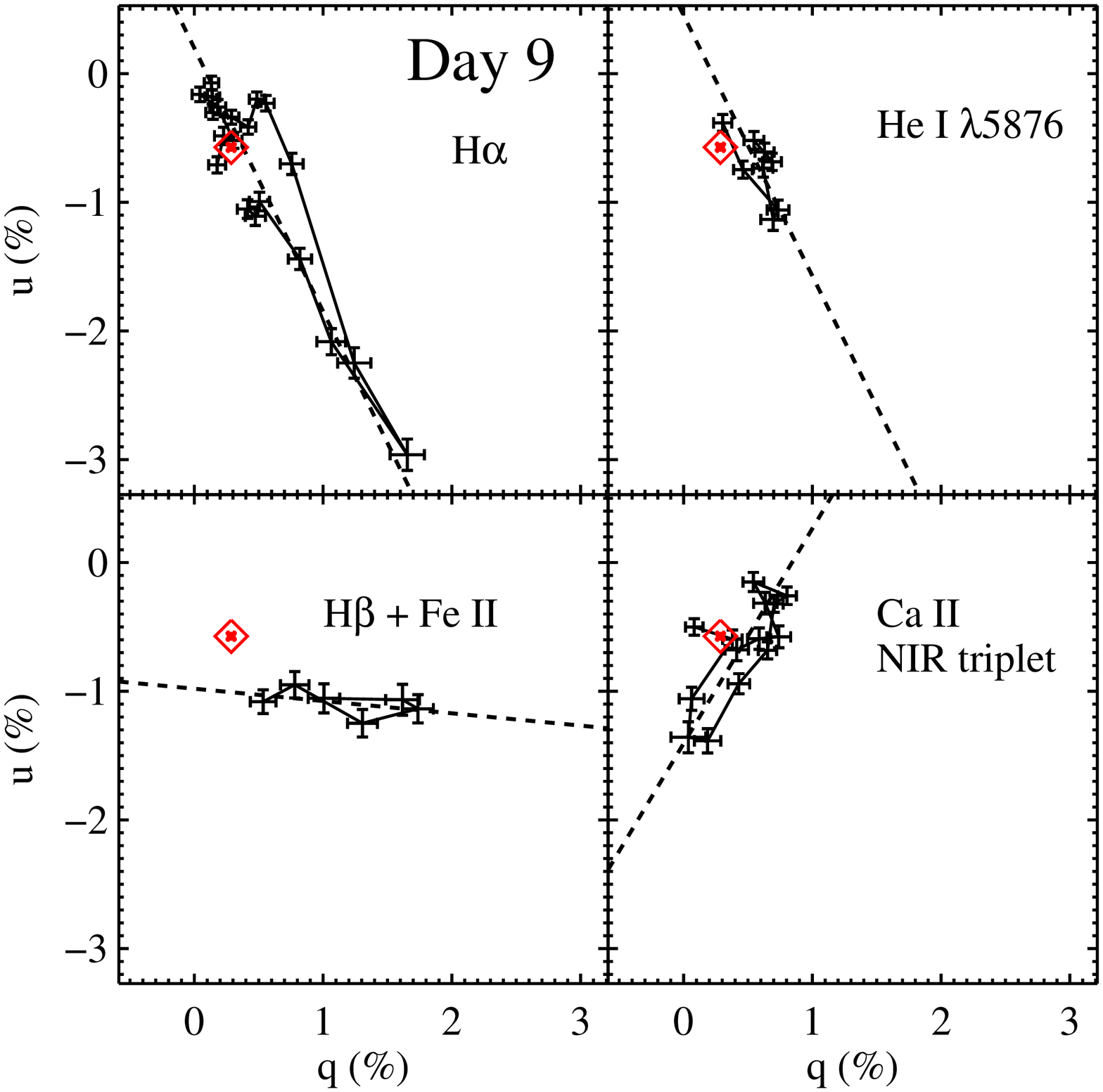}
\caption[Day 9 line-polarization data plotted in the $q-u$
  plane]{Line-polarization data from day 9 plotted in the $q-u$ plane,
  after subtraction of ISP.  The diamond in each panel (red in the
  online version) represents
  the continuum polarization measured from the relatively line-free
  7000$-$7450~\AA\ region (the error bar is smaller than the plot
  symbol).  The data points with error bars in black 
  in each panel represent wavelength bins corresponding to the
  indicated spectral features (\hbeta\ + \ion{Fe}{2}: 4600$-$4900~\AA;
  \ion{He}{1}: 5450$-$5850~\AA; \hal: 5950$-$6950~\AA; \ion{Ca}{2}:
  7900$-$9200~\AA).  The solid lines connect the points in order of
  wavelength to better show the loop structure.  The overplotted
  dashed lines represent linear fits to the line-polarization data to
  guide the eye.  Between
  days 6 and 9, the \hal\ and \ion{He}{1} $\lambda$5876 polarization
  angles changed little.  However, the polarization feature near
  H$\beta$ is now dominated by blending with \ion{Fe}{2} and has
  rotated to a different polarization angle reflecting the iron
  distribution.  The 
  \ion{Ca}{2} data on day 9 show a polarization angle that is clearly
  different from that of the Balmer and \ion{He}{1} lines. 
}
\label{ax_n2lineplot}
\end{figure}

Figures~\ref{ax_n1lineplot} and \ref{ax_n2lineplot} show 
wavelength intervals around important lines in the ISP-subtracted 
day 6 and 9 data in the $q-u$ plane. In addition to
presenting the data in the $q-u$ plane, it is helpful to plot the
ISP-subtracted spectropolarimetry versus wavelength.  To do that, we
will first want to choose more optimal coordinates than Stokes $q$ and
$u$, which are defined in reference to the position angle (P.A.) of
North on the sky, which is unrelated to any property of the supernova.
The strong line-polarization modulations seen in Figure~\ref{ax_rawpolplot}
appear as strong, almost linear features in the $q-u$ plane in
Figures \ref{ax_n1lineplot} and \ref{ax_n2lineplot}.  Despite
individually being almost linear in the $q-u$ plane, the line
polarization features collectively are not aligned and prefer
different polarization angles.  This is a signature of strong
deviations from axisymmetry and that the data are not well described
by a single P.A., so there is no unambiguous choice of a preferred
axis to use when rotating our coordinate system.

Inspection of Figure~\ref{ax_rawpolplot} reveals that the wavelength
interval of 7000$-$7450~\AA\ on day 9 is devoid of strong line
features in either the flux spectrum or the polarization.  This region
falls between the P-Cygni absorption troughs from \ion{He}{1}
$\lambda$7065 and \ion{O}{1} $\lambda$7774, and we regard it as
representative of the line-free continuum.  The integrated
ISP-subtracted polarization is $P = 0.64 \pm 0.02$\% at a P.A. of
$148\fdg3 \pm 1\fdg1$.  This point is plotted as a diamond (red in the
online version) in each
panel of Figure~\ref{ax_n2lineplot}.  Interestingly, it falls near the
dashed line showing a fit to the \hal\ data points.  The
ISP-subtracted polarization in the peak of the polarization feature
associated with the \hal\ absorption trough on day~9 (6190$-$6230~\AA)
is $P = 3.4 \pm 0.1$\% at an angle of $149\fdg5 \pm 1\fdg1$,
consistent with the polarization angle of our adopted 
continuum region.  The similar \hal\ polarization peak on day 6
(6120$-$6180~\AA) has a polarization of $2.8 \pm 0.2$\% at an angle
of $150\fdg9 \pm 1\fdg7$.

\begin{figure*}
\plotone{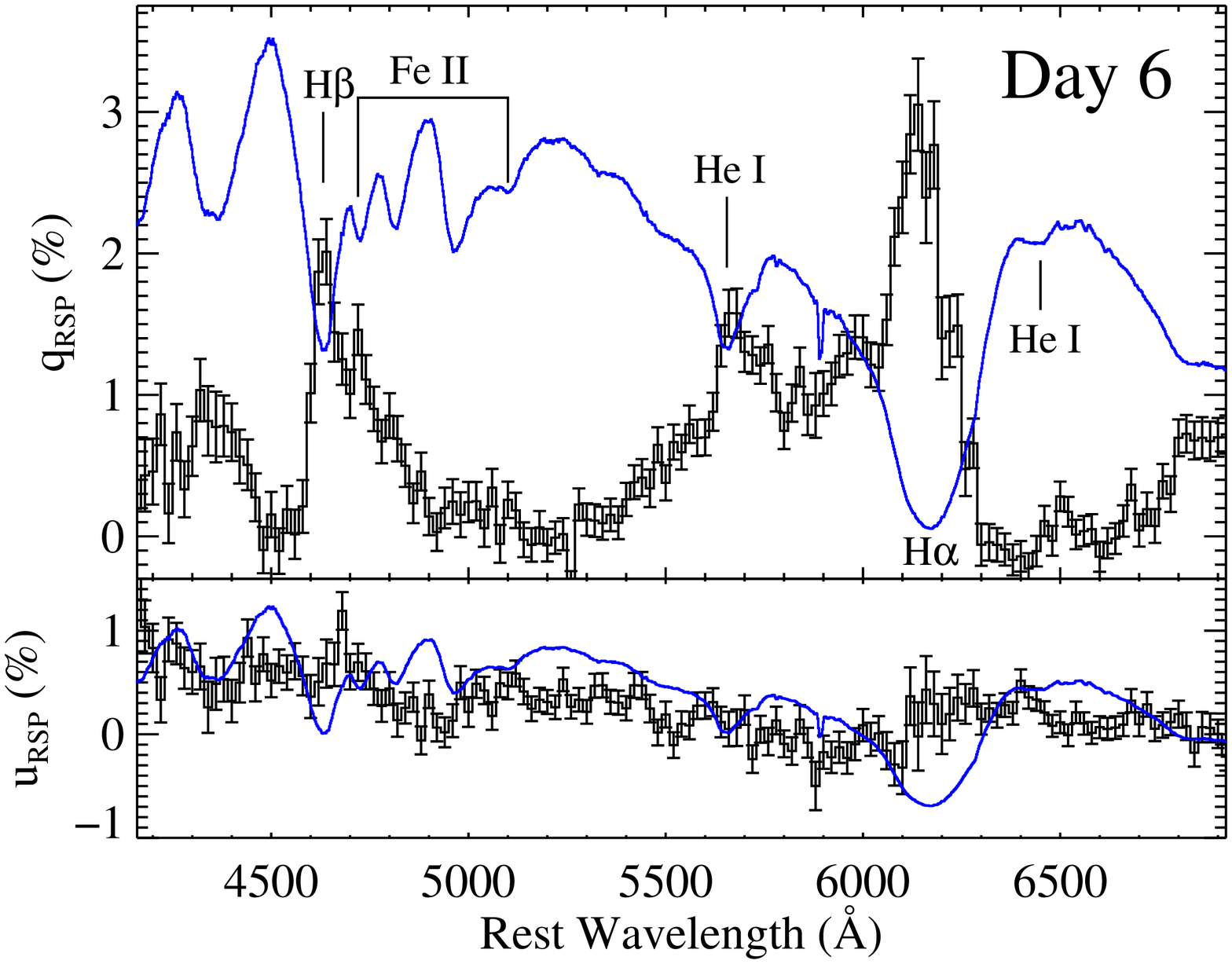}
\caption[Day 6 spectropolarimetry corrected for ISP.]{Day 6
  spectropolarimetry corrected for ISP and rotated to align
  \qrsp\ with the day 9 continuum polarization angle of $148\fdg3$,
  presented in 20~\AA\ bins.  Absorption minima of lines of
  polarimetric interest are labeled.
  If \ax\ were axisymmetric, \qrsp\ would be an estimator of the total
  polarization and \ursp\ would be centered on zero.  The deviations
  of \ursp\ from zero are a sign of deviations from axisymmetry.
  Error bars on this and all subsequent plots are 1$\sigma$.
}
\label{ax_day6rotplot}
\end{figure*}

\begin{figure*}
\plotone{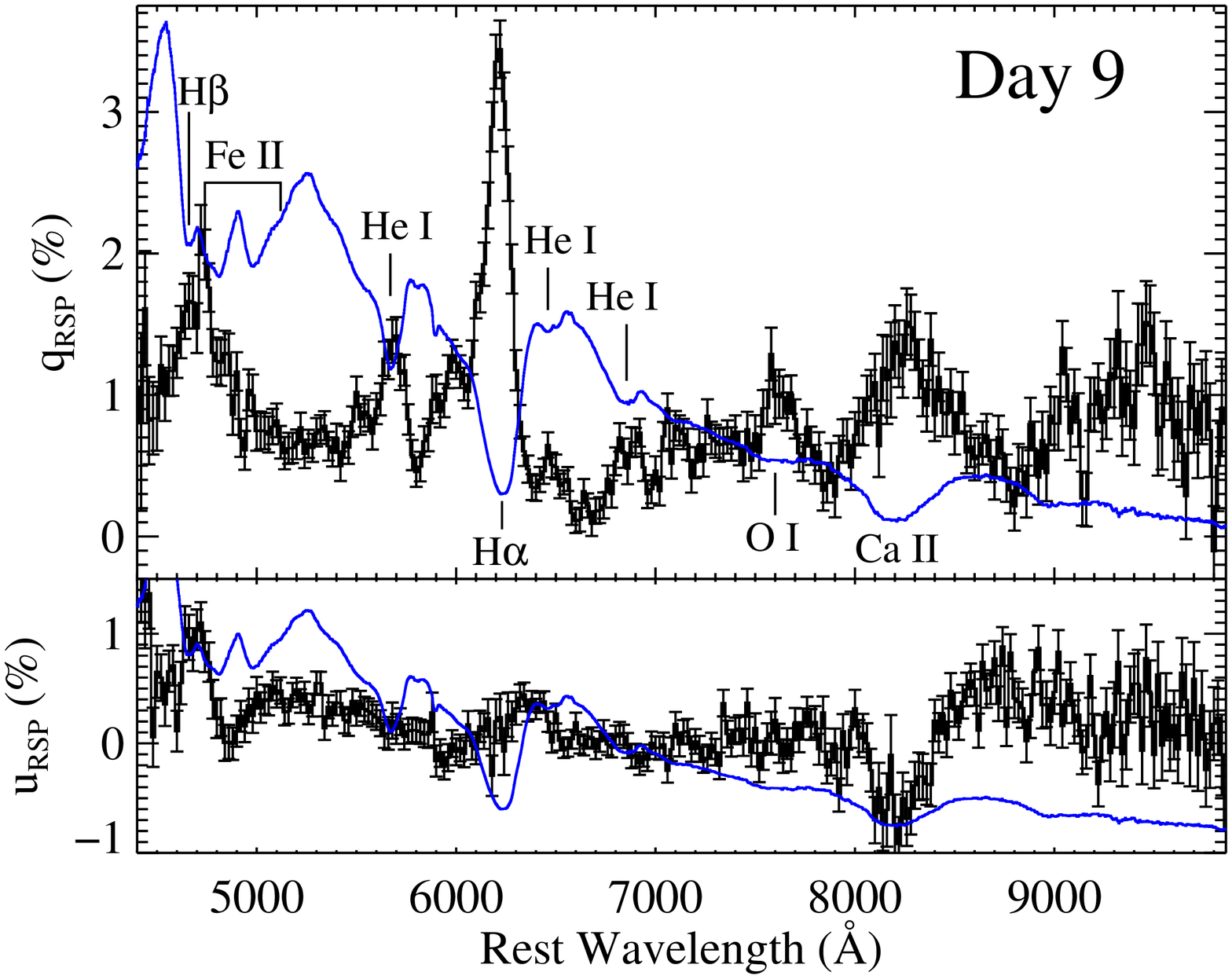}
\caption[Day 9 spectropolarimetry corrected for ISP.]{Day 9
  spectropolarimetry corrected for ISP and rotated to align
  \qrsp\ with the 7000$-$7450~\AA\ continuum polarization angle of
  $148\fdg3$, presented in 20~\AA\ bins.  Strong polarization
  modulations are seen in association with each P-Cygni feature.  The
  absorption minima of lines discussed in the text are labeled.
}
\label{ax_day9rotplot}
\end{figure*}

This excellent agreement between the continuum and \hal\ trough
polarization angles (along with the polarizations of several other
absorption lines, see below) gives us confidence that our value for
the ISP was accurately determined and that the continuum P.A.
($148\fdg3$) is physically significant in this object.  We will choose
a new coordinate system of rotated Stokes parameters (RSP; Trammell et
al. 1993a; Tran 1995) which aligns \qrsp\ with this P.A.  If the
ejecta were axisymmetric, \qrsp\ would be a good estimator of the
total polarization and \ursp\ would be near zero.  Deviations from
zero in \ursp\ are a signature of deviations from axisymmetry.  An
additional motivation for choosing RSP is to avoid the statistical
bias of the formal polarization ($P = (q^2 + u^2)^{\frac{1}{2}}$) to
positive values.
The rotated spectropolarimetry is plotted versus wavelength in
Figures~\ref{ax_day6rotplot} and \ref{ax_day9rotplot}.

\subsection{Line Polarization}

\subsubsection{\hal\ Polarization}

\begin{figure}
\plotone{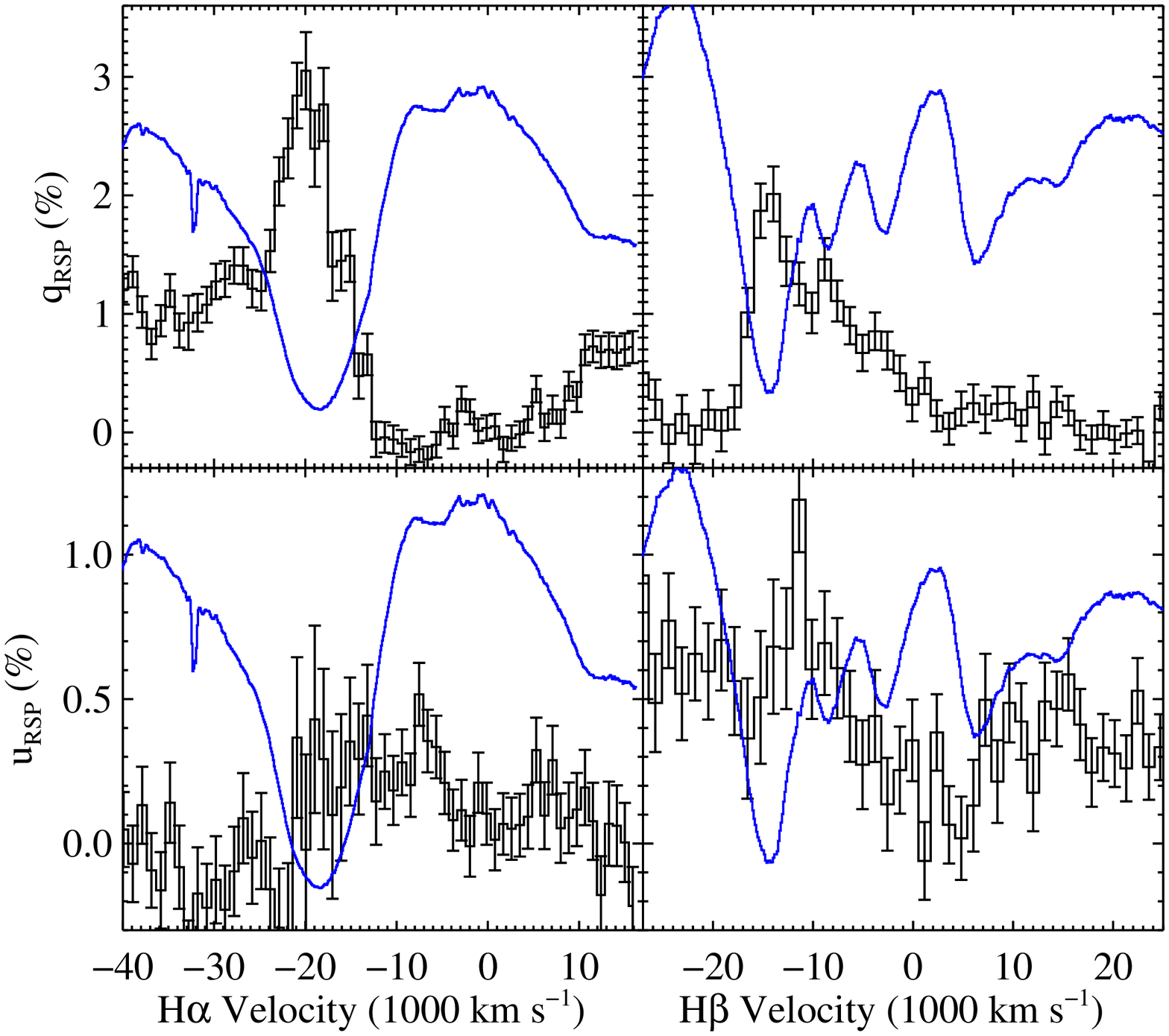}
\caption[Day 6 Balmer line polarization.]{\hal\ and \hbeta\ line
  polarization on day 6 versus velocity.  The gray line in each panel
  (blue in the online version) is the total-flux spectrum plotted to
  guide the eye.  The 
  absorption near $-$14,000 \kms\ in the panels on the right is due to
  \hbeta.  The notches redward of that are due to \ion{Fe}{2}
  multiplet 42.  The polarization peaks in \qrsp\ are associated with
  the absorption minima of \hal\ and \hbeta.
}
\label{ax_day6velplot}
\end{figure}

\begin{figure}
\plotone{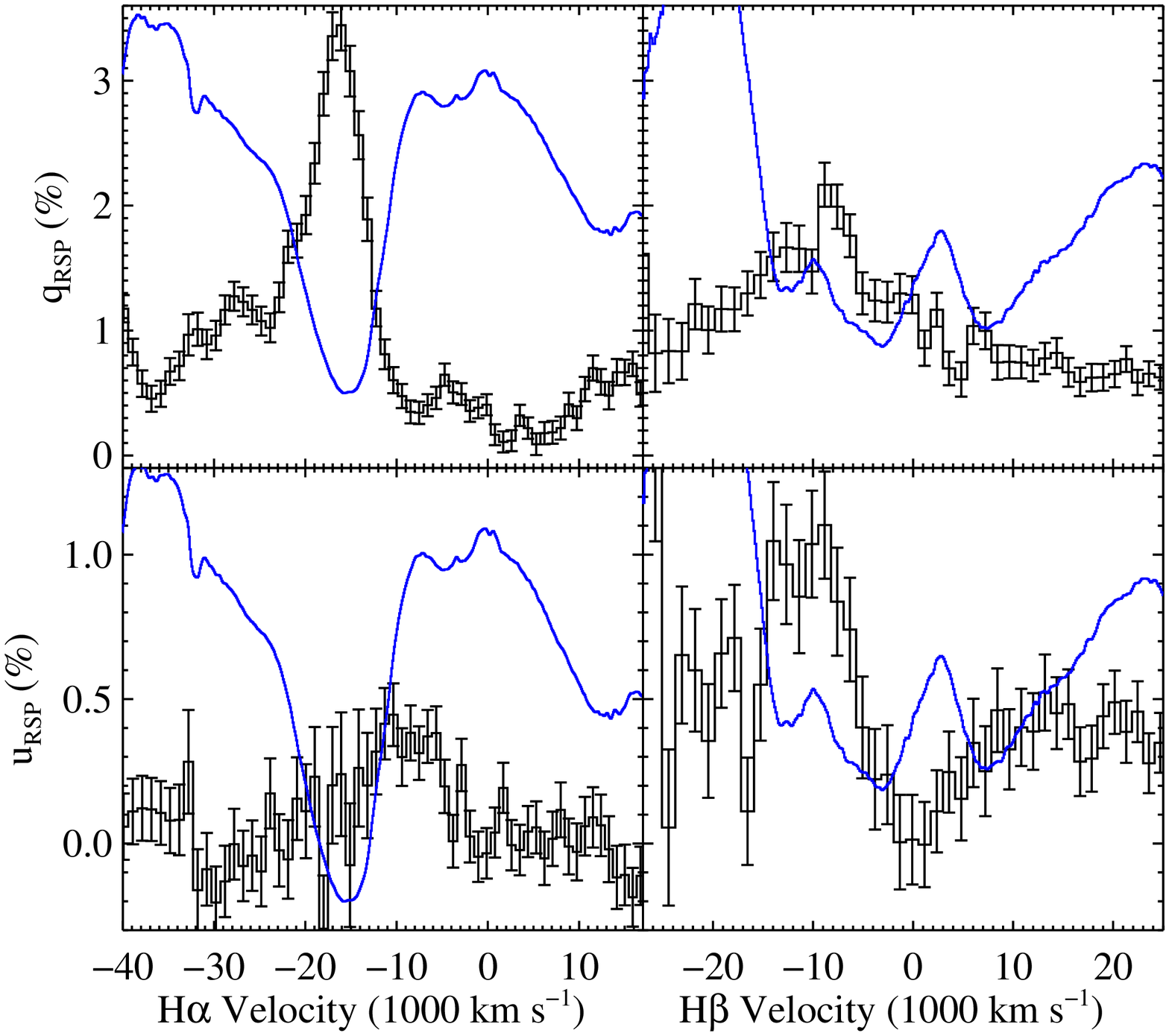}
\caption[Day 9 Balmer line polarization.]{\hal\ and \hbeta\ line
  polarization on day 9 versus velocity.  The gray line in each panel
  (blue in the online version) is the total-flux spectrum plotted to
  guide the eye.  \hbeta\ is 
  not prominent in the total-flux spectrum at this time (it is the
  absorption near $-$13,000 \kms\ in the panels on the right) and the
  \ion{Fe}{2} lines dominate the observed polarization signal.
}
\label{ax_day9velplot}
\end{figure}

We will start our exploration of the line-polarization behavior of
\ax\ with \hal\ because it is the strongest 
line in the spectrum and can be used as a reference.  In
addition to the full spectropolarimetry in Figures
\ref{ax_day6rotplot} and \ref{ax_day9rotplot}, we have plotted
\qrsp\ and \ursp\ versus velocity for \hal\ and \hbeta\ in Figures
\ref{ax_day6velplot} and \ref{ax_day9velplot}.

Ignoring for the moment the modulations in \ursp, the basic shapes of
the \hal\ polarization features on days 6 and 9 match the ``inverted''
P-Cygni polarization features discussed above. To test the
expectation of low polarization near the line peak at zero velocity
(or alternatively, our choice of ISP), we integrated the polarization
over  6600$-$6700~\AA\ to estimate the \hal\ peak polarization, while
avoiding most of the effects of the \ion{He}{1} $\lambda$6678
absorption \citep{maund01ig}.  We obtain values for (\qrsp, \ursp) of
($0.06\pm0.05$\%, $0.12\pm0.05$\%) and ($0.17\pm0.04$\%,
$0.03\pm0.04$\%) on days 6 and 9, respectively.  These low values for
the \hal\ peak polarization despite some contamination from
\ion{He}{1} $\lambda$6678 give us additional confidence that our
derived ISP is reasonable. 

Similar peaks in polarization can be seen in \qrsp\ associated with
most of the P-Cygni absorption features in Figures \ref{ax_day6rotplot}
and \ref{ax_day9rotplot}.  In particular, there are clear features at
\hbeta, the \ion{Fe}{2} lines near 5000~\AA, \ion{He}{1}
$\lambda$5876, \ion{O}{1} $\lambda$7774, and the \ion{Ca}{2} NIR
triplet, as well as weaker peaks due to \ion{He}{1} $\lambda$6678 and
$\lambda$7065.  In a purely axisymmetric system, all of the line
features would show the inverted P-Cygni features in \qrsp\ but no
modulation in \ursp, which is in clear conflict with the strong
features seen in \ursp\ for some lines in Figures \ref{ax_day6rotplot}
and \ref{ax_day9rotplot}.  This is not an artifact of a poor ISP
correction or coordinate rotation.  The $q-u$ line-polarization plots
in Figures~\ref{ax_n1lineplot} and \ref{ax_n2lineplot} show that
different lines have quite different polarization behaviors which are
not aligned and indicate that the geometry of the ejecta of
\ax\ was quite complex. 

In addition to the evidence for deviations from axisymmetry seen from
the ensemble of line features, individual lines show that the
spatial distribution or excitation of individual chemical elements is
complex.  For example, the \hal\ data points in the upper-left panel
of Figure~\ref{ax_n1lineplot} are connected in order of wavelength by a
line.  If the \hal\ geometry were axisymmetric, the points would fall
in a line.  However, the modulation seen in \ursp\ manifests itself in
this plot as a two-dimensional spread of the \hal\ points.  This
spread is not random, but shows a systematic trend with wavelength in
a manner consistent with the $q-u$ ``loops'' first identified by
\citet{cropper88} in spectropolarimetry of SN~1987A.  Loops are
not confined to \hal; all the lines plotted in
Figures \ref{ax_n1lineplot} and \ref{ax_n2lineplot} show evidence for
loop-like structure.

\citet{ch92} proposed that individual clumps of $^{56}$Ni could result
in \hal\ excitation asymmetries that produce wavelength-dependent
polarization angle rotations across the line profile.  While the loops
in SN~1987A were seen in \hal\ and \ion{Ca}{2}, subsequent work has
found them to be ubiquitous in core-collapse SNe \citep{ww08}.  The
interpretation of loops and their implied velocity-dependent
polarization angles has remained obscure.  One possibility is that the
clumpiness of the ejecta results in each isovelocity surface
intersecting a somewhat different spatial distribution of each
chemical element.  However, if the clumps only existed on small
scales, the polarization angles in different velocity bins would be
essentially independent.  The systematic rotation of the polarization
angle with velocity implies that the relevant asphericities are
coherent on larger scales.

\citet{ka03} investigated the loop seen in the \ion{Ca}{2} NIR
triplet of the SN Ia 2001el and found that loops were generically
produced by a line-scattering region with a misaligned axis of
symmetry relative to the continuum and therefore that inversion of the
observed Stokes parameters for a single object was highly
underconstrained.  Despite SN 2001el being a SN Ia, the models of
\citet{ka03} are surprisingly relevant to \ax, at least for \hal.
In SN 2001el, the geometry involved a high-velocity absorption from
\ion{Ca}{2} detached above the photosphere.  As discussed above, the
\hal\ line in \ax\ is also detached from the photosphere in a
high-velocity shell.  The misaligned aspherical shell models of
\citet{ka03} may therefore be quite relevant.  With the long axis of
the \hal\ loop being largely aligned with the continuum polarization,
the \hal\ distribution likely has only a slight misalignment with the
symmetry axis of the photosphere.

While the models of \citet{ka03} may be able to explain loops seen in
detached lines like \hal, the relevance of these models to lines
forming closer to the photosphere (such as \ion{He}{1}) is less
clear.  An additional difficulty is with the amplitude of the
polarization modulation.  A polarization of 3.4\% in the \hal\ line is
very high, particularly for a supernova being observed so soon after
explosion.  Similarly large polarizations ($P \lesssim 4$\%) were seen
in the trough of the \ion{Ca}{2} NIR triplet in SN 1987A
\citep{cropper88,je91,je91catalog}, but only at late times (day 100).
\citet{je91} was unable to successfully model such a large line
polarization.  He invoked a special excitation asymmetry, especially
to explain the differences between \hal\ and \ion{Ca}{2} at that
time.

\citet{maund05bf} found similarly high polarization values in the
\ion{Ca}{2} lines in the peculiar SN Ib/c 2005bf before maximum light
and \citet{maund08D} observed large ($P \approx 2$\%) calcium
polarizations in SN 2008D.  A common thread is that at the times of
large line polarization, the lines in question formed at high
velocities and were detached from the photosphere.  Evidently,
such a high-velocity, detached-shell geometry is quite favorable for
the production of significant line polarization.  That the lines
showing this effect are frequently due to calcium may just be due to
the relative strength of calcium lines in supernova ejecta
material at the low densities of most high-velocity material.

One last factor to mention in the context of very strong
line-polarization values is deviations of the geometry of the
high-velocity material from a spheroid.  Partial photospheric
obscuration at different velocity surfaces is responsible for the line
polarization seen from a detached aspherical shell \citep{ka03}, but
the polarizing effectiveness is limited by the spheroidal geometry.
If the absorbing material has a geometry far from that of a spheroid, 
such as a toroid or large clumps, then partial obscuration can very
effectively generate large polarizations \citep{ka03}.  Such
non-spheroidal geometries could also result in the appearance of the
\hal\ line being quite different in strength or velocity along other
lines of sight.

\subsubsection{Other Lines}

The polarization near \hbeta\ is more complicated than at \hal.  On
day 6, the locus of points in the $q-u$ plane
(Fig.~\ref{ax_n1lineplot}) is not linear or a simple loop.  The long
axis of the distribution is largely aligned with the
\hal\ orientation, but not exactly.  The peak polarization at the
absorption minimum, 
integrated over 4610$-$4650~\AA, is $P = 1.99 \pm 0.14$\% at a P.A.
of 157$\pm 2\degr$, which is slightly different from that at \hal.
The right-hand panels of Figure~\ref{ax_day6velplot} show that the
polarization peak is centered on the absorption minimum.  Unlike \hal,
\qrsp\ does not go to zero near zero velocity.  Instead,
\qrsp\ remains high at velocities redward of the absorption minimum.
Similarly, the polarization in \ursp\ is elevated above zero.  We
interpret this behavior as being due to polarization induced by the
superposed \ion{Fe}{2} lines of multiplet 42, which are visible as the
three notches in the total-flux spectrum redward of \hbeta\ in
Figure~\ref{ax_day6velplot}.

The spectrum of \ax\ was undergoing rapid evolution at this time.  By
our next epoch of spectropolarimetry, day 9, both the spectrum and
polarization near \hbeta\ had changed significantly.  The total-flux
spectrum in the right-hand panels of Figure~\ref{ax_day9velplot} shows
that the absorption from \hbeta\ had significantly weakened as the
photosphere had receded through the hydrogen envelope.  Meanwhile, the
neighboring \ion{Fe}{2} lines had strengthened.  The polarization data
points from this region are plotted in the lower-left panel of
Figure~\ref{ax_n2lineplot}.  The polarization feature clearly is no
longer oriented similarly to \hal\ and prefers another direction which
presumably reflects the iron distribution.  The polarization in the
core of the \hbeta\ absorption (integrated over 4630$-$4680~\AA) is
$P = 1.94 \pm 0.12$\% at a P.A. of 163$\pm 2\degr$, but the
polarization angle changes rapidly to the red as the effects of the
\ion{Fe}{2} lines manifest themselves.

The other lines shown in Figures~\ref{ax_n1lineplot} and
\ref{ax_n2lineplot} also exhibit loops.  The long axis of the \ion{He}{1}
$\lambda$5876 loop is largely aligned with the continuum and
\hal\ polarization axis, although the loop is responsible for some
deviation from that axis.  The amplitude of the modulation across
$\lambda$5876 is about 0.9\% on day 9.  The polarization peak on day 6
(5630$-$5700~\AA) has a polarization of $1.47 \pm 0.08$\% at a P.A. of
$150\fdg7 \pm 1\fdg6$, evolving on day 9 (5640$-$5740~\AA) to a
polarization of $1.30 \pm 0.06$\% at a P.A. of $151\fdg2
\pm 1\fdg3$.  The P.A. on both nights is consistent with the \hal\ and
continuum P.A.  The weaker modulations ($\Delta P 
\approx 0.3$\%) associated with \ion{He}{1} $\lambda$6678 and
$\lambda$7065 also appear to be mostly in the direction of \qrsp,
although the lower significance of those features prevents us from
studying any substructure. 

The loop formed by the \ion{Ca}{2} NIR triplet feature on day 9 is
clearly not aligned with any other ion in the spectrum or with the
continuum.  This result is independent of any ISP subtraction or
rotation of coordinates.  The relatively small wavelength dependence
of ISP means that ISP subtraction approximates a shift of the origin
of the $q-u$ plane.  The differing behaviors of \hal\ and the NIR
triplet can also be seen in the uncorrected $q$ and $u$
spectropolarimetry plotted in the right-hand panel of
Figure~\ref{ax_rawpolplot}.  While the \hal\ trough polarization
deviates from the continuum polarization in the positive $q$, negative
$u$ direction, the NIR triplet absorption deviates in the negative
$q$, negative $u$ direction.  The net polarization modulation, or the
length of the long axis of the loop, is about 1.3\%.

A small polarization increase is evident in the \ion{O}{1}
$\lambda$7774 absorption trough in Figure~\ref{ax_day9rotplot}, although
the noise is too high relative to the amplitude of the feature to
see a loop if one were present.  The integrated polarization over the
7550$-$7700~\AA\ interval is (\qrsp, \ursp) = (0.97\%, 0.03\%), with
error bars of 0.06\% on each Stokes parameter.  This means that the
polarization in the oxygen absorption is elevated by 0.33\% above the
continuum level, but has the same P.A.

In summary, the hydrogen, helium, and oxygen lines in \ax\ all have
largely consistent orientations (ignoring the small modulations in
\ursp\ due to the loops), which they share with the continuum at red
wavelengths.  It is natural to identify this angle with the symmetry
axis of the supernova ejecta.  Calcium and iron appear to have
different orientations.  \ax\ is not unique in this respect
\citep{maund05bf,maund08D}, which we will discuss further below.

\subsection{Late-Time Polarization}

\begin{figure}
\plotone{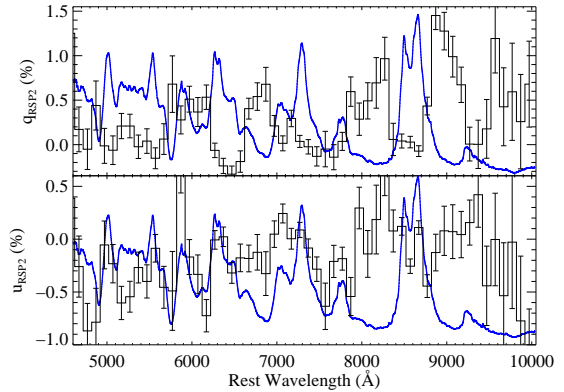}
\caption[Day 99 spectropolarimetry corrected for ISP.]{Day 99
  spectropolarimetry after correction for ISP, and presented in
  100~\AA\ bins.  The data have been rotated to a new coordinate
  system ($q_{\rm{RSP2}}$, $u_{\rm{RSP2}}$), chosen to align $q_{\rm{RSP2}}$ with the
  strong continuum polarization signal near 8000 and 9000~\AA.
}
\label{ax_laterotplot}
\end{figure}

We have also corrected our late-time spectropolarimetry for ISP, but
we have chosen a  new coordinate system to present the data.  Regions
of high continuum 
polarization are evident in Figure~\ref{ax_latepolplot} between the
nebular emission lines.  We chose two such regions on either side of
the \ion{Ca}{2} NIR triplet, 7850$-$8250~\AA\ and 8850$-$9150~\AA, and
integrated the ISP-subtracted Stokes parameters.  We obtained
polarization angles of $174 \pm 5\degr$ and $175 \pm 3\degr$,
respectively.  We chose to rotate our new coordinate system
($q_{\rm{RSP2}}$, $u_{\rm{RSP2}}$) to this preferred axis (weighted average of
174$\fdg$3) and plotted the data in Figure~\ref{ax_laterotplot}.  This
method places most of the highest continuum polarization peaks into
$q_{\rm{RSP2}}$.  If we had used the other rotated coordinate system,
\ursp\ would have shown large deviations from zero.

The basic pattern of high continuum polarization still holds in
$q_{\rm{RSP2}}$.  The region of highest polarization is our continuum
window of 8850$-$9150~\AA, which has a net ISP-corrected polarization
of $1.28 \pm 0.11$\%, indicative of large asphericities in the inner
core of the SN ejecta.  The polarization dips down to nearly zero at
the wavelengths of strong emission lines by construction due to our
choice for the ISP.

Curiously, the polarization at bluer wavelengths is nearly zero in
$q_{\rm{RSP2}}$, but is systematically negative in $u_{\rm{RSP2}}$.  The
integrated polarization over 5050$-$5950~\AA\ (approximating the $V$
band) is $P = 0.26 \pm 0.06$\% at a P.A. (prior to rotation)
of $143 \pm 5\degr$.  This P.A. is completely consistent with that
of the continuum polarization angle measured on day 9 and used to
rotate into our original ($q_{\rm{RSP}}$, $u_{\rm{RSP}}$) coordinates, but is
inconsistent with the P.A. in our redder continuum windows.

Usually the consistency of the continuum polarization angle from early
to late times would be interpreted as resulting from a single axis of
symmetry in the ejecta, but the rotation of the continuum
polarization angle with wavelength at late times shows that the inner
ejecta do not have a simple 
axisymmetric geometry.  While such a rotation is potentially a symptom
of poor ISP removal, the substantial scatter in the $q-u$ plane we
observe shows that no single ISP could align the wavelength-dependent
continuum angle.  Perhaps the continuum polarization angle at late
times is affected by individual clumps of $^{56}$Ni producing regions
of enhanced electron scattering \citep{ch92}.  The $V$-band
polarization angle could also be modified by the spatial distribution
of the iron lines mostly responsible for the pseudocontinuum at those
wavelengths. 

Lastly, we note that one potential alternative method for determining
the ISP would be to assume that the polarization at blue wavelengths
is nearly zero at late times under the hypothesis that the large
amount of \ion{Fe}{2} opacity should result in complete depolarization
of the continuum \citepeg{ho01}.  The low $V$-band polarization we do
measure after subtraction of our preferred ISP shows that this method
would result in only a slightly different ISP subtraction.

If we had used the late-time $V$-band polarization to measure the ISP
and then rotated our early-time data into a coordinate system aligned
with the day 9 continuum polarization, there would be a few negative
consequences.  Specifically, the \hal\ line polarizations in our new
\qrsp\ vector would be somewhat offset such that the polarization peaks
in the absorption troughs would still be at positive values of
\qrsp\ but the polarization at the line peaks would be at negative
values of \qrsp.  Such a polarization profile is certainly possible
\citepeg{jef89}, but we prefer the result of our chosen ISP making the
polarization of the emission-line peaks at late times and in the
\hal\ peaks at early times nearly zero.  A small residual  
polarization in the $V$ band at late times may indicate that the iron
responsible for the line opacity and depolarization of the continuum
might not be physically located exterior to the continuum formation
region and is instead cospatial with the scattering electrons such
that some polarized flux can escape to the distant observer.

\section{Discussion\label{ax_discuss}}

\subsection{Hydrogen Envelope}

The well-sampled spectroscopic sequences presented in \S3 and by P08
and \citet{taub11} demonstrate that \ax\ made a transition
from the hydrogen-dominated spectra of a SN II to the helium-dominated
spectra of a SN Ib, reflecting a progenitor that had lost most but not
all of its hydrogen envelope \citep{crockett08}.  A key question is
how much hydrogen the progenitor retained at the time of explosion.
The progenitor system of the prototypical SN IIb 1993J has been
directly detected \citep{aldering94,maund04,ms09} and determined to
have been a K-type supergiant interacting with a B-type supergiant
companion.  The direct observations of the progenitor of \ax\ are
still ambiguous due to possible contamination from other stars and
await final, late-time template images \citep{crockett08}.  However,
the progenitor of \ax\ was probably not similar to that of SN 
1993J, and was either an even more highly stripped star in an
interacting binary (superposed on a stellar cluster) or a WR star
with a small residual hydrogen envelope (a WNL/WNH star; Crockett et
al. 2008), although \citet{sc08} argue that the WNH stars 
are still undergoing core H burning.  The WR progenitor model may also
have difficulty in explaining the shape of the bolometric light curve
of \ax\ \citep{crockett08}.  \citet{cs09} have recently proposed that
\ax\ is one of a group of SNe~IIb coming from compact WR-like
progenitors, while SN~1993J is the exemplar of a different class of
SNe~IIb which appear to come from more extended progenitors.

\citet{ryder04} observed modulations in the radio light
curve of the SN IIb 2001ig which they attributed to the interaction
between the WR wind of the progenitor and a hot companion, a
conclusion reinforced when \citet{ryder06} found a supergiant
star at the location of the supernova long after explosion, which they
identified as the companion to the progenitor.  \citet{soderberg06}
instead preferred to interpret the variations in the radio light curve
as signs of a variable wind from a single massive WR star.  These
results are relevant to \ax\ because its radio light curve
\citep{roming09} shows similar modulations to those of SN 2001ig, although 
the pre-supernova mass-loss rate of the \ax\ progenitor appears to have
been lower than in other SNe IIb \citep{roming09}.

Spectroscopically, it seems clear that \ax\ had significantly less
hydrogen in its envelope than SN 1993J.  The Balmer lines in the
optical spectra of \ax\ disappeared at a much earlier epoch than in
SN~1993J (\S3; P08), while the NIR spectra never showed
much evidence for hydrogen.  Typical models for the progenitor of SN
1993J \citep{shigeyama94,woosley94} find a hydrogen-envelope mass of
0.2$-$0.9 M$_{\odot}$, with \citet{hf96} measuring
$\sim$0.3 M$_{\odot}$ of hydrogen directly from the late-time spectra.

\citet{elmhamdi06} obtained spectroscopic hydrogen-mass estimates for
several stripped-envelope SNe using some simple assumptions.
They derive 0.7 M$_{\odot}$ of hydrogen for SN 1993J (on the high
side of other estimates), but only 0.1 M$_{\odot}$ for the
SN IIb (or hydrogen-rich SN Ib) 2000H \citep{branchIb}, which had
stronger absorption from \hal\ at late times than
\ax\ (Fig. \ref{ax_heiplot}).  \citet{branchIb} found that masses of
order 10$^{-2}$ M$_{\odot}$ of hydrogen would be necessary in local
thermodynamic equilibrium (LTE) to produce an optical depth in
\hal\ of order unity near maximum light in SNe Ib.
\citet{tominaga05} estimated that $\sim0.02$ M$_{\odot}$ of hydrogen
at high velocities would reproduce the observed \hal\ absorption in
the SN Ib/c 2005bf, which, although peculiar and differing from \ax\ in
other ways, was the closest
spectroscopic match to \ax\ near maximum light.  The broad-lined SN
IIb 2003bg had stronger \hal\ absorption at late times than
\ax\ \citep{hamuy09}, which \citet{mazzali09} were able to model using
$\sim$0.05~M$_{\odot}$ of hydrogen.
Combined, these estimates indicate that \ax\ likely ejected $\sim$few
$\times$ $0.01$ M$_{\odot}$ of hydrogen, although a full non-LTE spectral
analysis is necessary to be more precise.  The non-LTE models of
\citet{james2010} found that the \hal\ lines of SNe~1999dn and 2000H
could be modeled using $\lesssim10^{-3}$~M$_{\odot}$ and
$\lesssim0.2$~M$_{\odot}$ of hydrogen in the outer envelope,
respectively.

\subsection{Spectropolarimetry Comparison}

\begin{figure}
\plotone{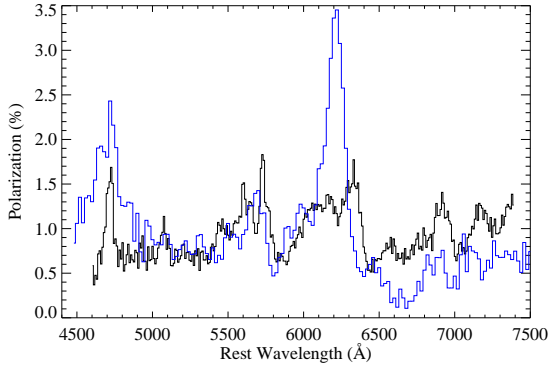}
\caption[Comparison with SN 1993J spectropolarimetry]{Day 9
  polarization data for \ax\ (gray, blue in the online version)
  compared to the SN 1993J data (black) from 1993 April 30 after
  correction for ISP \citep{tran97}. 
}
\label{ax_93jcompplot}
\end{figure}

SNe~IIb are surprisingly well represented in the published sample of
SN spectropolarimetry in comparison with SNe~IIP \citep{ww08}, given 
their relative rates.
Spectropolarimetry of SN~1993J was presented by \citet{thw93} and
\citet{tran97}, and the data of \citet{thw93} were subsequently
modeled in detail \citep{ho95,ho96}.  These observations were mostly
taken at early phases, when the photosphere was in the hydrogen-rich
outer envelope, while our \ax\ data were taken when the
photosphere was starting to recede through the helium core.
\citet{maund01ig} found that the spectropolarimetry of SN 2001ig
changed dramatically at a similar phase.  We have compared our day 9
spectropolarimetry to SN 1993J \citep{tran97} in
Figure~\ref{ax_93jcompplot}.  The epoch of the SN 1993J observations
was chosen to roughly match the strength of the \ion{He}{1}
lines with \ax\ so that the photosphere would be at an approximately
similar depth with respect to the helium-rich core.

The first thing to note from the comparison in
Figure~\ref{ax_93jcompplot} is that the polarization near 6600~\AA\ in
the peak of the \hal\ line is almost zero in \ax, as discussed above,
but is $\sim 0.7$\% in SN 1993J.  This reflects the different
assumptions used by \citet{tran97} to correct their SN 1993J data for
ISP.  Choosing a different ISP for SN 1993J could lower the 
overall level of polarization, including the amplitudes of the peaks
at the line features, and thus the inferred level of asphericity
\citep{ho96}.  The line features in the \ax\ data are already 
clearly stronger than in SN 1993J, particularly at \hal\ and the
\hbeta/\ion{Fe}{2} blend.

\citet{wa01} found that their spectra and spectropolarimetry of the SN 
IIb 1996cb closely resembled those of SN 1993J and proposed that the
objects had similar geometries oriented similarly with respect to our
line of sight.  They wondered if SNe IIb were a homogeneous subclass
and whether those objects would have appeared differently from a
different viewing angle.  \citet{lf05} presented a single epoch of
spectropolarimetry of the SN IIb 2003ed and found that it was also
broadly similar to SN 1993J.  The SN 2001ig data of \citet{maund01ig}
have some similarity to SN 1993J, but are more distinct than the other
objects.

To the extent that our spectropolarimetry of \ax\ resembles that of
SN~1993J, it may be a result of the same basic phenomenon 
(overlapping inverted P-Cygni line polarization profiles) in objects
with similar spectral features.  Thus, the similarity between the
spectropolarimetry of SNe 1993J and 1996cb noted by \citet{wa01} may
simply be a consequence of the close spectroscopic resemblance between
those two objects \citep{qiu99}.  SN 2001ig was quite
spectroscopically distinct from SN 1993J \citep{maund01ig,silverman09}
and had somewhat different polarimetry.  As shown in
\S3, \ax\ is the least spectroscopically similar of
these objects, and so the fact that it has the most distinct
polarimetry might have been expected. Furthermore, the speculation of
\citet{wa01} about the appearance of SNe IIb from different lines of
sight will be directly tested in one case due to the 
discovery of light echoes \citep{krause05,rest08} around the Galactic
SN IIb Cas~A \citep{krause08}, which have enabled the study of that
explosion from different vantage points \citep{rest10}.  If the high
\hal\ polarization we see in \ax\ is due to large clumps or a special
geometry, the appearance of the \hal\ line could be quite different
along other lines of sight.  Indeed, \citet{rest10} have found that
the \hal\ velocity of the Cas~A supernova was higher along one line
of sight compared to others that they studied.

Our analysis of the full spectral evolution of
\ax\ (\S3) found that the object was more similar to
normal SNe Ib than to SN 1993J, so it is of interest to compare to
the spectropolarimetry of those objects as well.  Unfortunately, SNe
Ib are not as well represented in the literature.  After the first
detection of possible intrinsic SN polarization in the SN Ib 1983N
\citep{mccall85}, the next SNe Ib observed spectropolarimetrically
were SNe 1997dq and 1998T, both of which had substantial ($P > 1$\%)
line-polarization features \citep{doug00}.  However, none of those
observations have been published.  The only comparison SNe Ib are
the unique SN Ib/c 2005bf \citep{maund05bf,tanaka05bf} and SN 2008D
\citep{maund08D}.

The SN Ib/c 2005bf had many unusual and unique photometric and
spectroscopic properties
\citep{tominaga05,folatelli06,parrent07,maeda07}.  However, our
automated comparisons using SNID in \S3 found it to be
the closest spectral match to \ax\ near maximum light.  Single epochs
of spectropolarimetry were obtained 6 days before \citep{maund05bf}
and 8 days after \citep{tanaka05bf} maximum light.
The early observations of \citet{maund05bf} show very large
polarizations in the troughs of the \ion{Ca}{2} H\&K and NIR triplet
absorptions ($P \lesssim 4$\%), with smaller ($P \lesssim 1$\%)
features present at the \ion{He}{1} $\lambda$5876 and \ion{Fe}{2}
lines.  The line features decreased in strength by the time of the
observations of \citet{tanaka05bf}, but at both epochs $q-u$ loops
were observed in the \ion{He}{1} and \ion{Ca}{2} lines.

One result common to the spectropolarimetry of SN 2005bf
\citep{maund05bf,tanaka05bf} and SN 2008D \citep{maund08D} is that the
line features did not share an axis of symmetry.  The loops observed
in almost all well-detected lines show that the individual features
have velocity-dependent angle rotations, but the overall range of
angles occupied by the lines is different for different species.  Even
the same lines observed at different epochs show rotations as the
photosphere recedes.  In particular, the \ion{He}{1}, \ion{Ca}{2}, and
\ion{Fe}{2} lines show misaligned polarization angles in both objects.
If the polarization feature near 4700~\AA\ in our day 9 data is mostly
due to \ion{Fe}{2}, then our \ax\ data show exactly the same
phenomenon (Fig.~\ref{ax_n2lineplot}).  Interestingly, our \ion{He}{1}
$\lambda$5876 data are largely aligned with both the continuum and
line-polarization angles of \hal\ and oxygen at that date.

The continuum polarization we measure on day 9 ($P = 0.64 \pm 0.02$\%)
is higher than that determined by \citet{maund08D} for SN 2008D ($P
\approx 0.2$\%) and \citet{maund05bf} for SN 2005bf at early times ($P
\approx 0.45$\%), although \citet{tanaka05bf} show that different ISP
assumptions for SN 2005bf could increase that value to $\sim1.2$\%.
In the context of the oblate spheroidal models of \citet{ho91}, our
\ax\ continuum polarization implies that the photosphere on day 9 has
a minimum axis ratio of about 1.15:1.  If we take the polarization in
our reddest continuum window on day 99 (1.28\%) as representative of
the late-time continuum photosphere, then the minimum asphericity
in the core rose to $\sim$35\%.

Also of interest to us is the evidence presented by \citet{maund05bf}
of only small polarization increases ($P \lesssim 0.2$\%) associated
with the high-velocity Balmer absorptions in SN 2005bf.  While our
early-time spectropolarimetry of \ax\ showed very large polarization
modulations across \hal, we were unable to obtain spectropolarimetry
near maximum light.  At that time, the \hal\ line was significantly
weaker and \ax\ resembled normal SNe Ib with a strong
6300~\AA\ feature.  An interesting question for future research is the
time evolution of polarization modulations at \hal\ in other
stripped-envelope SNe.  If the spectroscopic similarities between
\ax\ and SN 2005bf are meaningful, then perhaps future SNe Ib
observed near maximum light will also show weak \hal\ polarization
even if the line were strongly polarized at early times, due to the
decreased photospheric covering fraction or optical depth at late
times.  In this context, it is interesting to note that
\citet{maund08D} found a peak in the polarization ($P = 0.73 \pm
0.13$\%) of SN 2008D at 6260~\AA, which they 
observed was at a rather red wavelength to be associated with the
\ion{Si}{2} $\lambda$6355 absorption and tentatively suggested that
it instead may have been due to weak high-velocity \hal\ absorption at
$v \approx -$17,050 \kms.

\subsection{SN 2008ax as a Supernova Remnant}

The nebular-phase oxygen line profiles presented in \S3.4 are
asymmetric and some lines show multiple peaks.  In particular, the
double-peaked structure of the \ion{Mg}{1}] $\lambda$4571 emission
line is strong evidence against the possibility of a largely
blueshifted inner oxygen distribution, which was one possibility
explored by \citet{mili09}.  Instead, the multi-peaked line profiles
are likely a sign that models which invoke a clumpy inner oxygen
distribution are more correct.  Clumping due to Rayleigh-Taylor
instabilities in the explosion \citepeg{muller91} has been invoked in
the past to explain the multi-peaked line profiles observed in other
stripped-envelope SNe, including SN~1993J \citep{ma00}.  The
continuum polarization of 1.28\% observed in the early nebular phase
is also a sign that the inner ejecta of \ax\ were highly aspherical.
Spectral modeling is necessary to determine how much of the
polarization we measure at late times is due to global asphericities
versus clumping in the ejecta.

Our spectra of \ax\ as it entered the nebular phase show the strong
oxygen and calcium lines expected of an object that will develop into
an oxygen-rich supernova remnant, while our earliest optical spectra
demonstrated that a small amount of hydrogen was present in the outer
layers of the ejecta.  These properties are strongly reminiscent of
the Galactic supernova remnant Cas~A \citep{ck78,ck79}.
\citet{fesen87} identified fast-moving nitrogen-rich ejecta knots that
probably represent the outermost layers of the progenitor, which
apparently were composed of a small amount of hydrogen, much like a WN
star, and similar to one of the potential progenitor models for \ax.
The discovery by \citet{krause08} that the spectrum of the 
supernova explosion giving rise to Cas~A was similar to that of SN
1993J only strengthens the relevance of \ax\ to Cas~A.  Similarly,
\citet{tominaga05} identified Cas~A as a possible analog for SN
2005bf.  However, \citet{cs09} identified Cas~A as a member of the
SN~1993J-like SNe IIb with extended progenitors, while SN~2008ax was a
representative of their class of SNe IIb from compact progenitors.

Optical and X-ray studies of emission from the ejecta of Cas~A have
shown dramatic compositional inhomogeneities on all scales, from
small optical knots \citep{ck78} to the full remnant \citep{hwang04}.
Iron-rich ejecta from the core have penetrated and overturned the
silicon-rich layer that was situated above them in the progenitor
\citep{hughes00}, while not penetrating the overlying oxygen-rich
ejecta \citep{fesen06}.  The overall appearance of the remnant is very
aspherical and turbulent, with silicon-rich ``jets''
\citep{fesen01,hwang04}.

The lack of a simple, well-defined symmetry axis in our early-time
spectropolarimetry of \ax, along with the differing polarization
angles associated with different chemical elements and the
multi-peaked nebular line profiles, appears
unsurprising in the context of the complex morphology of Cas~A.
We note that some evolution after explosion has occurred in that
remnant due to deceleration by the surrounding gas \citep{fesen06},
and possibly also by energy input from radioactive decay
\citep{hughes00}.
Cas~A should give pause to anyone attempting to interpret the geometry
of distant SN explosions such as \ax\ in a smooth, quasi-spherical
or axially-symmetric fashion.  \citet{wheel08} have attempted to piece
together a coherent picture of the Cas~A observations in the context
of a jet-induced explosion.

\section{Conclusions}

This detailed study of \ax\ was made possible by the luck of a robotic
SN search taking an image of the nearby host galaxy only hours after
shock breakout from the SN \citep{mostardi08}, with the further luck
that the image obtained by another observer was able to bracket the
time of explosion so tightly \citep{arbour_cbet}.  This is reminiscent
of the fortuitous discovery of the X-ray burst accompanying SN 2008D
\citep{edo08}, which allowed numerous groups to obtain detailed
follow-up observations of that SN Ib starting soon after explosion
\citep{soderberg08,modjaz09,mazzali08,malesani09}.
Only with the early discovery of SNe can we hope to obtain crucial
spectral information about the outer atmospheres of their progenitor
stars.

Our first spectrum of \ax, taken only two days after shock
breakout, is one of the earliest spectra of a supernova ever obtained
and does not resemble any of the spectra of other core-collapse
SNe observed at such early times.  Strong and deep Balmer
absorptions were present, extending to very high velocities ($>$33,000
\kms).  We have presented an extensive sequence of optical and NIR
spectra of \ax\ in \S3.  Our optical spectra were taken with a
relatively high cadence during the first week after explosion and
capture the very rapid spectroscopic evolution at early times. 
The progenitor of \ax\ had only a very low-mass hydrogen envelope, so
the Balmer lines faded on the rise to maximum light and optical
\ion{He}{1} lines started to become prominent (see also P08).  This
spectroscopic transition defines SNe IIb.  Our NIR spectra, starting
only 10 days after shock breakout, showed little evidence of hydrogen
and were dominated by the helium lines expected in a SN Ib. 

\citet{branchIb} found that the spectra of many SNe Ib exhibited an
absorption near 6300~\AA\ and presented arguments in favor of an
identification with \hal.  Our early optical spectra of \ax\ show 
unambiguous Balmer lines, with the \hal\ absorption smoothly evolving
into an absorption near 6270~\AA.  By maximum light, the
optical-through-NIR spectra very closely resembled those of
the SN Ib~1999ex.  If the first spectrum of \ax\ had not been
taken until maximum light or shortly afterwards, as is typical of most
supernovae, it likely would have been classified as a
SN Ib with some hydrogen absorption, as were a few of our comparison
objects (Fig.~\ref{ax_heiplot}).

This suggests that the distinction between SNe IIb and Ib may in part
depend on the quality and timing of spectroscopic observations.
The case of SN~1993J was not ambiguous because that object had a
sufficiently thick hydrogen envelope for \hal\ to be clearly present
at all times.  At the other extreme, the SN Ib 2008D was observed very
shortly after explosion and never showed the strong Balmer absorptions
seen in \ax\ (Fig.~\ref{ax_earlyplot}).  However, most of the other
objects plotted in Figure~\ref{ax_heiplot} appear to be intermediate
between those two, and the distinction between those objects
spectroscopically classified as SNe Ib and as SNe IIb may not be
physically meaningful.  If the SNe Ib had been observed as soon
after explosion as \ax, then many of them likely would also have been
classified as SNe IIb.  Our NIR spectral sequence of
\ax\ (Fig.~\ref{ax_irplot}) and the radio light curves of
\citet{roming09} also support a closer correspondence of \ax\ to
typical SNe~Ib than to SN 1993J, despite the prominence of optical
hydrogen lines at early times.

These observations of \ax\ support the suggestion of \citet{branchIb}
that the spectra of objects classified as SNe Ib frequently exhibit
some \hal\ absorption.  Therefore, the progenitors of many SNe Ib must
explode with some
hydrogen in their outer envelopes.  The processes that 
consistently lead to such a small, but nonzero, amount of hydrogen in
the outer envelope before explosion are not understood, and represent
a challenge for stellar evolution models.  Resolution of the
ambiguities in the direct detection of the progenitor of
\ax\ \citep{crockett08}, along with studies of the progenitors of
future stripped-envelope SNe, will help clarify the situation. 

Our three epochs of spectropolarimetry and the nebular line profiles
demonstrate that \ax\ showed strong asphericities starting shortly
after explosion and continuing until the early nebular phase.  The
3.4\% polarization feature at \hal\ on day 9 was unusually strong and
may indicate that the appearance of the \hal\ absorption could depend
on the viewing orientation with respect to the object.
After correction for ISP, the hydrogen, helium, and oxygen
distributions were found to be largely aligned with the continuum axis
of symmetry, while the iron and calcium lines exhibited strong deviations
from this axial symmetry.  All lines with sufficiently good detections
also showed loops in the $q-u$ plane indicative of velocity-dependent
angle rotations of unknown origin.  These deviations from axisymmetry
appear to be common in stripped-envelope SNe, but their cause is
mysterious.  Detailed spectral synthesis, including polarization,
of explosion models is necessary to understand whether those models
are capable of reproducing these observations.

\acknowledgments

We acknowledge M. Ganeshalingam, C. V. Griffith, J. Hodge, K. Merriman,
R. Mostardi, D. Poznanski, and X. Wang for assistance with some of the
observations.  We would also like to thank the observers of the Lick
AGN Monitoring Project 2008 (LAMP) for obtaining some of these
\ax\ spectra during the nightly supernova observations, in particular
A. J. Barth, V. N. Bennert, M. C. Bentz, G. Canalizo, K. D. Hiner,
C. E. Thornton, J. L. Walsh, and J.-H. Woo. 
Some of the data presented herein were obtained at the W. M. Keck
Observatory, which is operated as a scientific partnership among the
California Institute of Technology, the University of California, and
the National Aeronautics and Space Administration; it was made
possible by the generous financial support of the W. M. Keck
Foundation.  The authors wish to recognize and acknowledge the very
significant cultural role and reverence that the summit of Mauna Kea
has always had within the indigenous Hawaiian community; we are most
fortunate to have the opportunity to conduct observations from this
mountain.  We also would like to thank the expert assistance of the
Keck and Lick staffs in making these observations possible.  Data for
this manuscript was obtained by Visiting Astronomers at the
Infrared Telescope Facility, which is operated by the University of
Hawaii under Cooperative Agreement no. NNX-08AE38A with the National
Aeronautics and Space Administration, Science Mission Directorate,
Planetary Astronomy Program.  We thank Alan Tokunaga and the IRTF
staff for their help performing timely observations. A.V.F.'s
supernova research group at UC Berkeley has been supported by NSF
grants  AST-0607485 and AST-0908886, as well as by the TABASGO
Foundation. M.M. is supported by a fellowship from the Miller
Institute for Basic Research in Science.  D.K.L., R.J.R., and
R.W.R. were supported by The Aerospace Corporation's Independent
Research and Development program. 

{\it Facilities:} \facility{Shane (Kast)}, \facility{Keck:I (LRIS,
  HIRES)}, \facility{Keck:II (ESI)}, \facility{IRTF (SpeX)}

\appendix

High-resolution spectroscopy of SNe can also be used as a tool
to study the ISM of SN host galaxies.  Our HIRES spectra of
\ax\ reveal absorptions from a number of diffuse interstellar bands
(DIBs; Herbig 1975), as 
well as the \ion{Na}{1} and \ion{K}{1} lines shown in
Figure~\ref{ax_naiplot}.  The absorbers responsible for DIBs are still
largely unknown \citepeg{herbig95}, so determining the presence and
relative strengths of the absorptions in environments far removed from
the local conditions of the ISM in the Galaxy may provide significant
clues.  The smooth continua of bright SNe provide a good
backdrop to measure the otherwise hard-to-study properties of DIBs in
external galaxies.  DIB features have been measured in a handful of
SNe so far (e.g., Rich 1987; Sollerman et al. 2005; Cox \& Patat
2008).  Most of the objects which have been studied to date are SNe
Ia.  A core-collapse SN such as \ax\ has the potential to probe
different ISM conditions due to its likely proximity to the sites of
massive-star formation.

\begin{figure}
\includegraphics[width=3.in]{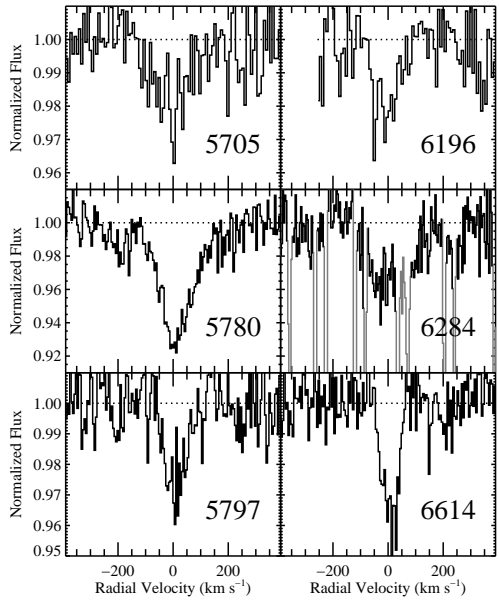}
\caption{DIBs in NGC 4490.  Each panel is labeled with the wavelength
  of the DIB feature.  The $\lambda$6284 DIB is heavily contaminated
  by telluric absorption.  The narrow telluric absorption features in
  that panel are grayed out and there is still significant
  uncertainty in the continuum level.
}
\label{dibsplot}
\end{figure}

An in-depth study of DIB features in NGC 4490 is beyond the scope of
this work.  However, we have plotted some prominent DIBs in
Figure~\ref{dibsplot} and measured their EWs in Table~\ref{dibtab} for
the benefit of other researchers.  The measured EWs for the atomic
lines are integrated over $\pm$40~\kms around our adopted value for
the recession velocity of \ax\ of 630~\kms.  The EWs of the DIBs are
integrated over $\pm$200~\kms.  The DIB near $\lambda$5705 is actually
a blend of two features at 5704.75~\AA\ and 5705.13~\AA\ in the
compilation of \citet{jd94}, but we have made no effort to decompose
the observed absorption.  Strong telluric absorption severely hampered
continuum placement around $\lambda$6284 and made any measurement of
the EW unreliable, so we have not reported a value.  The EWs of the
DIBs seen in an average Milky Way cloud with the same \ebv\ value as
\ax\ \citep{jd94} are also listed in Table~\ref{dibtab}.  The
$\lambda$6614 absorption seems relatively weak compared to the average
Galactic line of sight, an effect that was also seen along the line of
sight to SN 2006X in M100 by \citet{cp08}.

\begin{deluxetable}{lcrr|r}
\tablecaption{ISM Absorption Features in NGC 4490}
\tablehead{\colhead{Absorber} & \colhead{Wavelength} &
 \colhead{EW} & \colhead{Uncertainty\tablenotemark{a}} & 
 \colhead{MW\tablenotemark{b}} \\
 & (\AA) & (m\AA) & (m\AA) & (m\AA) }
\startdata
DIB & 5705 & 61.8 & 8.4 & 88.5 \\
DIB & 5780 & 228.0 & 7.0 & 289.5 \\
DIB & 5797 & 44.6 & 6.5 & 66.0 \\
DIB & 6196 & 52.1 & 10.3 & 30.5 \\
DIB & 6614 & 53.3 & 7.6 & 115.5 \\
\ion{Na}{1} & 5890 & 959.4 & 1.9 & \nodata \\
\ion{Na}{1} & 5896 & 827.4 & 2.0 & \nodata \\
\ion{K}{1} & 7699 & 142.0 & 6.2 & \nodata \\
\enddata
\tablenotetext{a}{Not including uncertainty in continuum placement.}
\tablenotetext{b}{Average value for a Galactic line of sight from
  \citet{jd94}, scaled to our value of \ebv = 0.5~mag.}
\label{dibtab}
\end{deluxetable}

\end{document}